\documentclass[reprint, aps, prl, superscriptaddress, longbibliography]{revtex4-2}
\usepackage{amsmath}
\usepackage{amssymb}
\usepackage{booktabs}
\usepackage{siunitx}
\usepackage{graphicx}
\graphicspath{{./Fig/}}

\begin{document}
\title{Transverse Polarization Gradient Entangling Gates for Trapped-Ion Quantum Computation}

\author{Jin-Ming Cui}
\thanks{These authors contributed equally to this work.}
\affiliation{Laboratory of Quantum Information, University of Science and Technology
of China, Hefei 230026, China }
\affiliation{Anhui Province Key Laboratory of Quantum Network, University of Science
and Technology of China, Hefei 230026, China}
\affiliation{ CAS Center For Excellence in Quantum Information and Quantum Physics,
University of Science and Technology of China, Hefei 230026, China }
\affiliation{ Hefei National Laboratory, University of Science and Technology of
China, Hefei 230088, China }

\author{Yan Chen}
\thanks{These authors contributed equally to this work.}
\affiliation{Laboratory of Quantum Information, University of Science and Technology
of China, Hefei 230026, China }
\affiliation{Anhui Province Key Laboratory of Quantum Network, University of Science
and Technology of China, Hefei 230026, China}
\affiliation{ CAS Center For Excellence in Quantum Information and Quantum Physics,
University of Science and Technology of China, Hefei 230026, China }

\author{Yi-Fan Zhou}
\affiliation{Laboratory of Quantum Information, University of Science and Technology
of China, Hefei 230026, China }
\affiliation{Anhui Province Key Laboratory of Quantum Network, University of Science
and Technology of China, Hefei 230026, China}
\affiliation{ CAS Center For Excellence in Quantum Information and Quantum Physics,
University of Science and Technology of China, Hefei 230026, China }

\author{Quan Long}
\affiliation{Laboratory of Quantum Information, University of Science and Technology
of China, Hefei 230026, China }
\affiliation{Anhui Province Key Laboratory of Quantum Network, University of Science
and Technology of China, Hefei 230026, China}
\affiliation{ CAS Center For Excellence in Quantum Information and Quantum Physics,
University of Science and Technology of China, Hefei 230026, China }

\author{En-Teng An}
\affiliation{Laboratory of Quantum Information, University of Science and Technology
of China, Hefei 230026, China }
\affiliation{Anhui Province Key Laboratory of Quantum Network, University of Science
and Technology of China, Hefei 230026, China}
\affiliation{ CAS Center For Excellence in Quantum Information and Quantum Physics,
University of Science and Technology of China, Hefei 230026, China }

\author{Ran He}
\affiliation{Laboratory of Quantum Information, University of Science and Technology
of China, Hefei 230026, China }
\affiliation{ CAS Center For Excellence in Quantum Information and Quantum Physics,
University of Science and Technology of China, Hefei 230026, China }

\author{Yun-Feng Huang}
\email[Corresponding author: ]{hyf@ustc.edu.cn}
\affiliation{Laboratory of Quantum Information, University of Science and Technology
of China, Hefei 230026, China }
\affiliation{Anhui Province Key Laboratory of Quantum Network, University of Science
and Technology of China, Hefei 230026, China}
\affiliation{ CAS Center For Excellence in Quantum Information and Quantum Physics,
University of Science and Technology of China, Hefei 230026, China }
\affiliation{ Hefei National Laboratory, University of Science and Technology of
China, Hefei 230088, China }

\author{Chuan-Feng Li}
\email[Corresponding author: ]{cfli@ustc.edu.cn}
\affiliation{Laboratory of Quantum Information, University of Science and Technology
of China, Hefei 230026, China }
\affiliation{Anhui Province Key Laboratory of Quantum Network, University of Science
and Technology of China, Hefei 230026, China}
\affiliation{ CAS Center For Excellence in Quantum Information and Quantum Physics,
University of Science and Technology of China, Hefei 230026, China }
\affiliation{ Hefei National Laboratory, University of Science and Technology of
China, Hefei 230088, China }

\author{Guang-Can Guo}
\affiliation{Laboratory of Quantum Information, University of Science and Technology
of China, Hefei 230026, China }
\affiliation{Anhui Province Key Laboratory of Quantum Network, University of Science
and Technology of China, Hefei 230026, China}
\affiliation{ CAS Center For Excellence in Quantum Information and Quantum Physics,
University of Science and Technology of China, Hefei 230026, China }
\affiliation{ Hefei National Laboratory, University of Science and Technology of
China, Hefei 230088, China }

\date{\today}
\begin{abstract}
The construction of entangling gates with individual addressing capability represents a crucial approach for implementing quantum computation in trapped ion crystals. Conventional entangling gate schemes typically rely on laser beam wave vectors to couple the ions' spin and motional degrees of freedom. Here, we experimentally demonstrate an alternative method that employs a polarization gradient field generated by a tightly focused laser beam—an approach theoretically proposed as the Magnus effect for quantum logic gate design [\emph{Phys. Rev. Res.} \textbf{5}, 033036 (2023)]. Using this technique, we perform Raman operations on nuclear spin qubits encoded in $^{171}\mathrm{Yb}^{+}$ ions, generating spin-dependent forces along axial motional modes in a linear trap.
By utilizing an acousto-optic deflector to create arbitrary spot pairs for individual ion addressing in two-ion (four-ion) chains, we achieve Mølmer-Sørensen gates with fidelities exceeding 98.5\% (97.2\%). Further improvements in numerical aperture (NA) and laser power could reduce gate durations while enhancing fidelity by orders of magnitude.
This method is compatible with—and can significantly simplify—optical tweezer gate proposals, where motional mode engineering enables scalable trapped-ion quantum computation. The technique can be readily extended to two-dimensional ion crystals, representing a key advancement toward large-scale trapped-ion quantum processors.
\end{abstract}
\maketitle

\emph{Introduction}--
Trapped-ion systems represent a leading platform for quantum computation,
offering long coherence times \cite{wang2017single,wang2021single,ruster2016long}, 
high-fidelity gate operations \cite{ballance2016high, gaebler2016high, cai2023entangling, loschnauer2024scalable, smith2025PRL}, 
and reconfigurable qubit connectivity \cite{pino2021demonstration, feng2023continuous}. 
Architectures based on trapped-ion crystals have emerged as a promising approach for building quantum processors,
with hundreds of ions successfully trapped and quantum simulations performed in two-dimensional crystals \cite{GuoDuan2024N,kiesenhofer2023controlling,shankar2022simulating}. 
Beyond quantum simulation demonstrations, a key requirement for quantum computation in this architecture
is the implementation of entangling gates with individual addressing (IA) capability.
Optical IA enables not only site-resolved qubit control but also facilitates
parallel quantum gate operations—essential for quantum error correction
and large-scale quantum algorithms \citep{LuKim2019N,FiggattMonroe2019N}, 
making optically addressed entangling gates a critical component of this approach.

Although entangling gates have been demonstrated in one-dimensional chains of tens of ions \cite{landsman2019two,chen2024benchmarking} and 
two-dimensional arrays of four ions \citep{HouDuan2024NC}, 
scaling quantum computation to hundreds of qubits in such systems remains challenging due to 
the difficulty of optimizing gates with large, dense motional modes in extended ion crystals.
To address this challenge, optical tweezer-based trapped ion quantum computation has been proposed as a promising approach \cite{olsacher2020scalable,shen2020NJP, mazzanti2021PRL, teoh2021manipulating, schwerdt2024scalable}. 
However, combining optical tweezer (OT) gates with IA-based entangling gates presents 
significant experimental challenges, as it requires multiple high-NA objective lenses along different axes. 
This limitation occurs because current IA methods couple to the laser propagation direction, 
while OT gates primarily modify motional modes perpendicular to the propagation axis—making 
single-lens OT gate implementations currently unfeasible.

\begin{figure*}
\centering
\includegraphics[width=\textwidth]{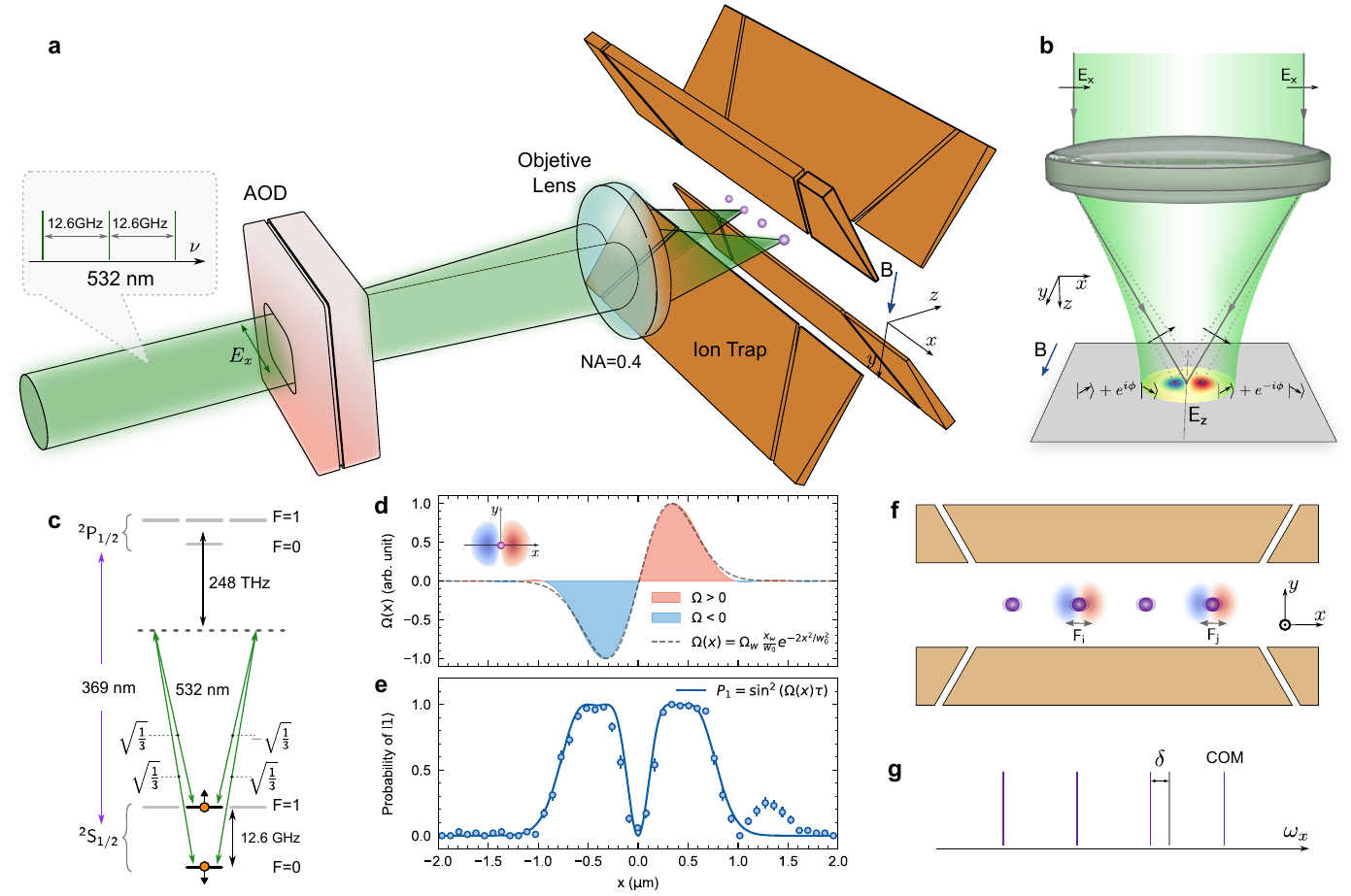}
\caption{\label{fig:1}
Principle of polarization-gradient-based individual addressing.
\textbf{a}, An acousto-optic deflector (AOD) is employed in the addressing system to generate multiple laser spots after the objective lens with high numerical aperture (NA). A multi-tone laser is used to couple to the Raman transition.
\textbf{b}, $x$-polarized-focused Gaussian beam creates longitudinal $E_z$ and elliptical polarization near focus. 
\textbf{c}, For $^{171}\mathrm{Yb}^{+}$, $\Omega(x)\propto\Omega_{+}^2(x)-\Omega_{-}^2(x)$, with $\Omega_{+/-}$ from right/left-circular components. 
\textbf{d},  Simulated Rabi profile (shaded) for NA=0.4, fit to $\Omega_w(x/w_0)e^{-2x^2/w_0^2}$ (dashed), showing maximal gradient at center. 
\textbf{e},  Spin-up probability measured by scanning the focus across the ion with $\tau=7\,\mu\mathrm{s}$ pulses, from which the Rabi profile in \textbf{d} is extracted via fitting.
\textbf{f}, Multi-qubit gates are implemented using aligned laser spots to address ions, coupling to an axial motional mode \textbf{g} to realize Mølmer-Sørensen gates.
}
\end{figure*}
In this work, we experimentally demonstrate Mølmer-Sørensen (MS) gates with a different IA coupling method,
which based on a polarization gradient generated by a tightly focused optical beam. 
This method enables coupling to motional modes perpendicular to the laser propagation direction, providing flexible IA capabilities—including entangling gates with axial modes in linear traps—while maintaining the simplified optical layout advantages of requiring only a single beam and high numerical aperture (NA) objective.
Using this technique, we perform carrier suppressed
Raman operations on nuclear spin qubits encoded in $^{171}\mathrm{Yb}^{+}$
ions and implement spin-dependent forces for entangling gates. By
dynamically steering the laser beam via an acousto-optic deflector,
we realize IA based entangling gates on two-ion and four-ion chains, achieving 
entangling gates with fidelities exceeding 98.5\% and 97.2\%, respectively.

Our method provides a promising path toward scalable trapped-ion quantum
computation. Its compatibility with optical tweezer architectures
and potential extension to two-dimensional ion arrays make it a versatile
tool for implementing parallel, high-fidelity gate operations in fault-tolerant
quantum computing schemes~\citep{ChengGong2024PRA,SpreeuwSpreeuw2020PRL,MazzantiSafaviNaini2023PRR}. 

\emph{Principle and Experiment}--
The interaction Hamiltonian between an optical field and a trapped ion qubit (optical or nuclear spin encoded) under the rotating wave approximation is~\citep{LeibfriedWineland2003RoMP}:
\begin{equation}
H_{I,k}=\frac{\hbar}{2}\Omega_{0}e^{i(\mathbf{k}\cdot\hat{\mathbf{x}}-\delta t+\phi)}\sigma_{+} + \text{H.c.},
\end{equation}
where $\Omega_{0}$ is the Rabi frequency, $\mathbf{k}$ the laser wave vector, $\hat{\mathbf{x}}$ the ion position operator (coupling to motional states), $\sigma_{+}$ the spin raising operator, and $\delta$, $\phi$ the laser detuning and phase respectively. 
For spin-motion coupling, $\mathbf{k}$ must be non-zero. In a Raman scheme driving nuclear spin qubits with two beams, the effective wave vector $\mathbf{k}=\mathbf{k}_{1}-\mathbf{k}_{2}$ requires $|\mathbf{k}_{1}|\approx|\mathbf{k}_{2}|$ with different propagation directions to ensure $\mathbf{k}\neq0$.

For one-dimensional motion along $x$ under Lamb-Dicke approximation:
\begin{equation}
H_{I,k}\approx\frac{\hbar}{2}\Omega_{0}\left[1+i\eta_k(a+a^\dagger)\right]\sigma_{+}e^{-i\delta t+i\phi} + \text{H.c.},
\end{equation}
where $\eta_k = k_x x_\omega$ is the Lamb-Dicke parameter with $x_\omega = \sqrt{\hbar/(2m\omega)}$ the ground state wave packet size. This wave vector coupling inherently maintains a dominant carrier transition ($\eta_k < 1$), limiting sideband operation speeds due to off-resonant carrier excitation.

Constructing a position-dependent Rabi frequency $\Omega_{0}\rightarrow\Omega(x)$ (in the laser spot's local coordinate system) enables more flexible spin-motion coupling. 
Notably, this approach allows spin-motion coupling even when $\mathbf{k} \approx 0$ (with co-propagating Raman beams):
\begin{equation}
H_{I,\Omega}=\frac{\hbar}{2}\Omega(x_{0}+\hat{x})\sigma_{+}e^{-i\delta t+i\phi}+\text{H.c.},
\end{equation}
where $x_{0}$ represents the ion's equilibrium position. When the motion amplitude is much smaller than $\Omega(x)$'s spatial profile, the Hamiltonian expands as:
\begin{align}
H_{I,\Omega} & \approx\frac{\hbar}{2}\left[\Omega(x_{0})+\Omega'(x_{0})\hat{x}\right]\sigma_{+}e^{-i\delta t+i\phi}+\text{H.c.}\nonumber \\
 & =\frac{\hbar}{2}\left[\Omega(x_{0})+\Omega_{s}(a+a^{\dagger})\right]\sigma_{+}e^{-i\delta t+i\phi}+\text{H.c.},\label{eq:Hamiltonian}
\end{align}
where $\Omega_{s}=\Omega'(x_{0})x_{\omega}$ denotes the sideband Rabi frequency and $\Omega'(x_{0})$ the Rabi gradient. 
The $\Omega(x)$ profile originates from tightly focused optical fields (individual addressing spots). Effective spin-motion coupling requires large Rabi gradients, while selecting $\Omega(x_{0})\rightarrow0$ suppresses carrier transitions. 
This carrier suppression, 
previously demonstrated in standing wave fields without individual addressing capability \cite{SanerBallance2023PRL}, 
now combines with IA for faster entanglement gates. 
The system's lateral motion coupling also enables an optical Magnus effect analogy \citep{SpreeuwSpreeuw2020PRL}, suggesting promising quantum logic gate implementations \citep{MazzantiSafaviNaini2023PRR}.

By engineering structured light field entering the focusing lens-- the spatial mode and polarization of the laser beam \cite{nape2022NP,wang2020AQS}
, the profile of $\Omega(x)$ can be controlled.
One approach to generating this profile is by utilizing the polarization
gradient of the tightly focused addressing spot. As shown in Fig.~\ref{fig:1}b, 
when linearly polarized ($x$-direction) light passes through a high-NA lens, a longitudinal $E_z$ component emerges at the focal plane.
If we approximate the $E_{z}$ component as resulting from combining two side paths through the lens in ray optics within the diffraction volume, the amplitude and phase of the longitudinal polarization clearly exhibit position dependence, creating a polarization gradient with elliptical components at the laser focus.

For a qubit encoded in the hyperfine clock transition of $^{2}S_{1/2}$ in $^{171}\mathrm{Yb}^{+}$ 
with quantization axis magnetic field $B$ along the $y$-axis, 
the Raman laser configuration shown in Fig. \ref{fig:1}c generates 
a Rabi profile (Fig.~\ref{fig:1}d) approximated by:
\begin{equation}
\Omega(x)=\Omega_{w}\frac{x}{w_{0}}e^{-2x^{2}/w_{0}^{2}},
\label{Eq:Profile}
\end{equation}
where $w_{0}\propto\lambda/\text{NA}$ is the beam waist. At spot center ($x=0$), 
maximum Rabi frequencies occur at $x_{m}=\pm w_{0}/2$ with $\Omega_{\mathrm{m}}=\pm\Omega_{w}/(2\sqrt{e})$, 
where $\Omega'(x_{m})=0$ decouples motion and spin. Positioning the ion at $x_{m}$ via laser spot shifting enables single-qubit rotations.

Our experiments utilize a linear chain of $^{171}\mathrm{Yb}^{+}$ 
ions confined in a segmented blade trap \citep{HeGuo2021RSI}. The 
qubit is encoded in the hyperfine clock states $|0\rangle\equiv|F=0,m_F=0\rangle$ 
and $|1\rangle\equiv|F=1,m_F=0\rangle$ of the $^2S_{1/2}$ 
manifold, forming a magnetic-field-insensitive qubit with a 12.64~GHz 
splitting. The quantization magnetic field (7.94~Gauss) aligns 
along the $z$-axis. The axial center-of-mass 
(COM) phonon mode frequency measures 287~kHz via DC 
tickling. After electromagnetically-induced transparency (EIT) cooling, 
the axial modes reach 5-7 phonons thermal population. 
Cooling, initialization, and state detection use 
369~nm lasers as described in \citep{OlmschenkMonroe2007PRA}. 
Fluorescence collected by a 0.64 NA objective lens and detected by a multi-channel PMT,
achieves state-preparation-and-measurement (SPAM) fidelity  $>$99.3\% .

\begin{figure}[htbp]
\centering 
\includegraphics{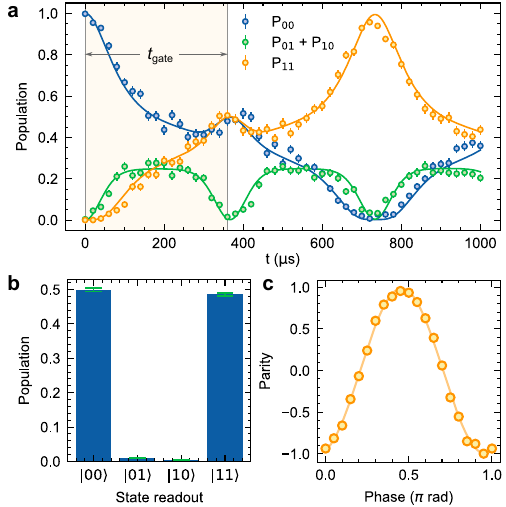} 
\caption{ \label{fig:fig3}
Mølmer-Sørensen entangling gate in a two-ion chain. 
\textbf{a}, Population evolution during gate operation, showing maximal entanglement at $t_{\mathrm{gate}}=367\,\mu\mathrm{s}$. Solid lines indicate theoretical predictions. 
\textbf{b}, Bell state population $P_{00}+P_{11}=0.985(1)$, averaged over 10,200 repetitions. 
\textbf{c}, Parity oscillation with 0.9887(1) contrast after applying $\pi/2$ pulses with variable phase $\phi$ to each ion. All data represent 400-repetition averages (unless noted), with error bars showing 1 standard error. State preparation and measurement (SPAM) errors are corrected \cite{shen2012correcting}.
}
\end{figure}

The experimental setup uses a multi-tone 532~nm Raman laser propagating
along the $z$-axis with linear polarization along the $x$-axis,
enabling the drive of axial motional modes. The Raman beam is generated
through a two-step modulation process: First, a 1064 nm seed laser
is phase-modulated using an electro-optic modulator (EOM) to generate
sidebands at 13.04 GHz. Subsequently, an unequal-arm Mach-Zehnder
interferometer constructed with a pair of acousto-optic modulators
(AOMs) driven at 200 MHz is used \footnote{See Supplemental Material at {[}URL{]} for details of the laser system, details of ion-laser alignment}. 
The $+1^{\mathrm{st}}$ and $-1^{\mathrm{st}}$ order diffracted beams produce a Raman beam with 12.64~GHz intensity modulation \citep{LiGuo2022OE,ChenGuo2024PRA}. 
The spatial Rabi frequency variation under a 0.4 NA objective lens follows Eq.~\ref{Eq:Profile}. 
Fig.~\ref{fig:1}e shows state $|1\rangle$ population evolution under fixed-duration ($\tau=7~\mu$s) Raman pulses versus beam position,
with the fitted curve giving $\Omega_{\mathrm{m}}=2\pi\times75.3$~kHz and $w_0=0.813(6)~\mu$m.
Ion motion introduces residual excitation ($P_1=0.05$) at beam center, corresponding to $\beta=\Omega_{\mathrm{m}}/\Omega(0)\approx15$ carrier suppression.
This ratio improves with lower ion temperatures and higher phonon frequencies.

To demonstrate the entangling gate, we employ a two-dimensional acousto-optic deflector (AOD) for beam steering, generating multi-site addressing spots. 
Each addressing spot is aligned to its corresponding ion using the measurement method shown in Fig.~\ref{fig:1}e, 
where scanning the beam and curve fitting determine each ion's precise position relative to the spot center, 
achieving alignment accuracy below 100 nm \cite{Note1}. 
As shown in Fig.~\ref{fig:1}f and Fig.~\ref{fig:1}g , the configured polarization gradient addressing couples to the axial motional modes, enabling MS gate implementation. 
Leveraging the spin-motion coupling from Eq.~\ref{eq:Hamiltonian}, MS gates are implemented through selective addressing of ions $i$ and $j$ as 
\[
U=\exp(-i\frac{\pi}{4}\sigma_i^x\otimes\sigma_j^x).
\]

\begin{figure}[htbp]
\centering 
\includegraphics{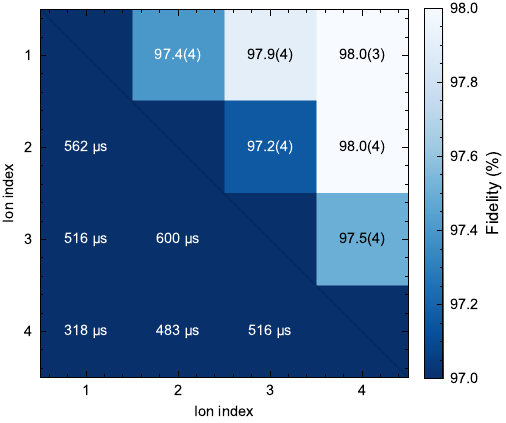} 
\caption{
Gate time and fidelity of MS gate for arbitrary two ions in the four-ion chain. 
The lower-left corner represents the gate time, while the upper-right
corner represents the Bell state fidelity. 
The population and paritydata are presented in the supplementary materials.
}
\label{fig:fig4}
\end{figure}

First, we demonstrated our scheme on a two-ion chain. After cooling
and initialization, a MS type gate is implemented
by simultaneously applying red-detuned and blue-detuned Raman laser
tones, coupling with the axial stretch mode. Fig. \ref{fig:fig3}a shows the evolution of spin populations as a function of interaction time. 
The fidelity of the gate is characterized through
parity oscillations and population measurements. Parity is measured
by applying a $\pi/2$ pulse to each ion after the MS gate, and is
defined as: 
$\mathrm{Parity}=P_{00}+P_{11}-P_{10}-P_{01},$
where $P_{ij}$ represents the population of the state $|ij\rangle$.
The parity oscillates as a function of the phase of the $\pi/2$ pulse,
and the contrast is defined as the oscillation amplitude. Fig. \ref{fig:fig3}c
demonstrates a contrast of $98.87(1)\%$ and $P_{00}+P_{11}=98.5(1)\%$,
leading to a gate fidelity calculated as
$F=(P_{00}+P_{11}+\text{Contrast})/{2}=98.7(1)\%.$

We have analyzed the sources of errors and presented them in Table \ref{table1}. In the current experiment, the dominant source of infidelity originates
from calibration errors in the red and blue sideband frequencies.
Due to the relatively low Rabi frequency, a detuning error of approximately
30 Hz can lead to an infidelity of 1\%. The presence of the optical tweezer
effects for changing the motional mode frequencies, 
and the differential AC Stark shifts for the clock states, 
causing energy levels changing along with the uncontrolled position drift between the laser focuses and ions, 
which limits more precise calibration, particularly for the
sidebands frequencies. The second leading contribution comes from
motional decoherence. For the stretch mode, with a blue sideband coherence
time exceeding 100~ms, the calculated infidelity induced by motional
decoherence is less than 0.25\%. The implementation of far-detuned
532 nm Raman laser and carrier suppression mechanisms in the protocol
effectively minimizes non-resonant effects and photon scattering contributions negligible. 
The gate fidelity can be further improved by simply
reducing the gate time and employing modulated drive waveforms \cite{shapira2023robust,shapira2018robust,wang2020high,kang2021batch,grzesiak2020efficient}.

\begin{table}[h]
\centering
\caption{Error budget for two qubit gate.}
\begin{tabular}{@{}lc@{}}
\toprule
\toprule
Error Source                 & Simulated Value  \\ \midrule
Mis-set Detuning             & \num{1.0e-2} \\
Motional Dephasing           & \num{2.4e-3} \\
Heating                      & \num{6.4e-4} \\
Rabi Frequency Fluctuation   & \num{2.5e-4} \\
Carrier Excitation       & \num{7.5e-6} \\
Thermal Errors           & \num{4.7e-6} \\ \midrule
Total Error              & \num{1.3e-2} \\ 
Experimental error       & \num{1.3(1)e-2}       \\ 
\bottomrule
\bottomrule
\end{tabular}
\label{table1}
\end{table}

To demonstrate scalability, we implement an MS gate on a four-ion
chain. As shown in Fig.~\ref{fig:fig4}, this implementation achieves a fidelity exceeding 97\% for arbitrary pairs of ions. 
The decrease in fidelity is primarily due to the extended
gate duration and crosstalk caused by intermodulation products in
the AOD. The gate duration could be shortened by increasing laser
power and using a larger NA objective.
AOD-induced crosstalk can be further suppressed by adding compensation
signals with opposite phases to counteract the intermodulation components \cite{shizhen2023optical}.

\emph{Conclusion and outlook}--
Our method to construct entangling gates offers several advantages: 
1) Flexible individual control, enabling axial motional mode coupling
as demonstrated; 
2) Carrier-suppressed coupling permitting large sideband Rabi frequencies
for fast gates; 
3) Robustness against initial thermal motion, since the equivalent Lamb-Dicke parameter $x_{\omega}/w_{0}$ is significantly smaller than in standing-wave configurations; 
4) Simplified optical setup requiring only a single beam and high-NA objective; 
5) Enhanced Raman phase stability from shared optical paths for all frequency components. 
Importantly, these advantages enable our scheme's extension to optical tweezer architectures with trapped ions, offering opportunities to overcome scalability bottlenecks in large ion crystal quantum computation.

The quantum gate fidelity is currently limited by both the operation time and motional state coherence time. The Raman sideband Rabi frequency scales approximately as $\mathrm{NA}^{4}$, indicating that increasing the numerical aperture (NA) of the addressing objective could substantially accelerate gate operations. An alternative approach involves enhancing the laser power. However, in our current setup, the Raman laser induces charging effects on the vacuum chamber's glass window, resulting in ion position drift that limits the usable laser power. This limitation can be mitigated by applying an indium tin oxide (ITO) coating to the inner surface of the glass window. Additionally, employing structured light modes such as hollow-core Laguerre-Gaussian beams suppresses differential AC shifts between clock states, thereby improving the overall fidelity.

\begin{acknowledgments}
	We thanks Robert J. C. Spreeuw in University of Amsterdam for fruitful discussions. This work was supported by the National Key Research and Development Program of China (Grant No. 2024YFA1409403), 
    the National Natural Science Foundation of China (Grant  No. 11734015, and No. 12204455), 
    the Innovation Program for Quantum Science and Technology (Grant No. 2021ZD0301604 and No. 2021ZD0301200), 
    and the Key Research Program of Frontier Sciences, CAS (Grant No. QYZDY-SSWSLH003).
	
	J.-M. Cui and Y. Chen contributed equally to this work.
\end{acknowledgments}

\bibliography{axial-mode-gate}

\begin{thebibliography}{42}%
\makeatletter
\providecommand \@ifxundefined [1]{%
 \@ifx{#1\undefined}
}%
\providecommand \@ifnum [1]{%
 \ifnum #1\expandafter \@firstoftwo
 \else \expandafter \@secondoftwo
 \fi
}%
\providecommand \@ifx [1]{%
 \ifx #1\expandafter \@firstoftwo
 \else \expandafter \@secondoftwo
 \fi
}%
\providecommand \natexlab [1]{#1}%
\providecommand \enquote  [1]{``#1''}%
\providecommand \bibnamefont  [1]{#1}%
\providecommand \bibfnamefont [1]{#1}%
\providecommand \citenamefont [1]{#1}%
\providecommand \href@noop [0]{\@secondoftwo}%
\providecommand \href [0]{\begingroup \@sanitize@url \@href}%
\providecommand \@href[1]{\@@startlink{#1}\@@href}%
\providecommand \@@href[1]{\endgroup#1\@@endlink}%
\providecommand \@sanitize@url [0]{\catcode `\\12\catcode `\$12\catcode
  `\&12\catcode `\#12\catcode `\^12\catcode `\_12\catcode `\%12\relax}%
\providecommand \@@startlink[1]{}%
\providecommand \@@endlink[0]{}%
\providecommand \url  [0]{\begingroup\@sanitize@url \@url }%
\providecommand \@url [1]{\endgroup\@href {#1}{\urlprefix }}%
\providecommand \urlprefix  [0]{URL }%
\providecommand \Eprint [0]{\href }%
\providecommand \doibase [0]{https://doi.org/}%
\providecommand \selectlanguage [0]{\@gobble}%
\providecommand \bibinfo  [0]{\@secondoftwo}%
\providecommand \bibfield  [0]{\@secondoftwo}%
\providecommand \translation [1]{[#1]}%
\providecommand \BibitemOpen [0]{}%
\providecommand \bibitemStop [0]{}%
\providecommand \bibitemNoStop [0]{.\EOS\space}%
\providecommand \EOS [0]{\spacefactor3000\relax}%
\providecommand \BibitemShut  [1]{\csname bibitem#1\endcsname}%
\let\auto@bib@innerbib\@empty
\bibitem [{\citenamefont {Wang}\ \emph {et~al.}(2017)\citenamefont {Wang},
  \citenamefont {Um}, \citenamefont {Zhang}, \citenamefont {An}, \citenamefont
  {Lyu}, \citenamefont {Zhang}, \citenamefont {Duan}, \citenamefont {Yum},\
  and\ \citenamefont {Kim}}]{wang2017single}%
  \BibitemOpen
  \bibfield  {author} {\bibinfo {author} {\bibfnamefont {Y.}~\bibnamefont
  {Wang}}, \bibinfo {author} {\bibfnamefont {M.}~\bibnamefont {Um}}, \bibinfo
  {author} {\bibfnamefont {J.}~\bibnamefont {Zhang}}, \bibinfo {author}
  {\bibfnamefont {S.}~\bibnamefont {An}}, \bibinfo {author} {\bibfnamefont
  {M.}~\bibnamefont {Lyu}}, \bibinfo {author} {\bibfnamefont {J.-N.}\
  \bibnamefont {Zhang}}, \bibinfo {author} {\bibfnamefont {L.-M.}\ \bibnamefont
  {Duan}}, \bibinfo {author} {\bibfnamefont {D.}~\bibnamefont {Yum}},\ and\
  \bibinfo {author} {\bibfnamefont {K.}~\bibnamefont {Kim}},\ }\bibfield
  {title} {\bibinfo {title} {Single-qubit quantum memory exceeding ten-minute
  coherence time},\ }\href@noop {} {\bibfield  {journal} {\bibinfo  {journal}
  {Nature Photonics}\ }\textbf {\bibinfo {volume} {11}},\ \bibinfo {pages}
  {646} (\bibinfo {year} {2017})}\BibitemShut {NoStop}%
\bibitem [{\citenamefont {Wang}\ \emph {et~al.}(2021)\citenamefont {Wang},
  \citenamefont {Luan}, \citenamefont {Qiao}, \citenamefont {Um}, \citenamefont
  {Zhang}, \citenamefont {Wang}, \citenamefont {Yuan}, \citenamefont {Gu},
  \citenamefont {Zhang},\ and\ \citenamefont {Kim}}]{wang2021single}%
  \BibitemOpen
  \bibfield  {author} {\bibinfo {author} {\bibfnamefont {P.}~\bibnamefont
  {Wang}}, \bibinfo {author} {\bibfnamefont {C.-Y.}\ \bibnamefont {Luan}},
  \bibinfo {author} {\bibfnamefont {M.}~\bibnamefont {Qiao}}, \bibinfo {author}
  {\bibfnamefont {M.}~\bibnamefont {Um}}, \bibinfo {author} {\bibfnamefont
  {J.}~\bibnamefont {Zhang}}, \bibinfo {author} {\bibfnamefont
  {Y.}~\bibnamefont {Wang}}, \bibinfo {author} {\bibfnamefont {X.}~\bibnamefont
  {Yuan}}, \bibinfo {author} {\bibfnamefont {M.}~\bibnamefont {Gu}}, \bibinfo
  {author} {\bibfnamefont {J.}~\bibnamefont {Zhang}},\ and\ \bibinfo {author}
  {\bibfnamefont {K.}~\bibnamefont {Kim}},\ }\bibfield  {title} {\bibinfo
  {title} {Single ion qubit with estimated coherence time exceeding one hour},\
  }\href@noop {} {\bibfield  {journal} {\bibinfo  {journal} {Nature
  communications}\ }\textbf {\bibinfo {volume} {12}},\ \bibinfo {pages} {233}
  (\bibinfo {year} {2021})}\BibitemShut {NoStop}%
\bibitem [{\citenamefont {Ruster}\ \emph {et~al.}(2016)\citenamefont {Ruster},
  \citenamefont {Schmiegelow}, \citenamefont {Kaufmann}, \citenamefont
  {Warschburger}, \citenamefont {Schmidt-Kaler},\ and\ \citenamefont
  {Poschinger}}]{ruster2016long}%
  \BibitemOpen
  \bibfield  {author} {\bibinfo {author} {\bibfnamefont {T.}~\bibnamefont
  {Ruster}}, \bibinfo {author} {\bibfnamefont {C.~T.}\ \bibnamefont
  {Schmiegelow}}, \bibinfo {author} {\bibfnamefont {H.}~\bibnamefont
  {Kaufmann}}, \bibinfo {author} {\bibfnamefont {C.}~\bibnamefont
  {Warschburger}}, \bibinfo {author} {\bibfnamefont {F.}~\bibnamefont
  {Schmidt-Kaler}},\ and\ \bibinfo {author} {\bibfnamefont {U.~G.}\
  \bibnamefont {Poschinger}},\ }\bibfield  {title} {\bibinfo {title} {A
  long-lived zeeman trapped-ion qubit},\ }\href@noop {} {\bibfield  {journal}
  {\bibinfo  {journal} {Applied Physics B}\ }\textbf {\bibinfo {volume}
  {122}},\ \bibinfo {pages} {254} (\bibinfo {year} {2016})}\BibitemShut
  {NoStop}%
\bibitem [{\citenamefont {Ballance}\ \emph {et~al.}(2016)\citenamefont
  {Ballance}, \citenamefont {Harty}, \citenamefont {Linke}, \citenamefont
  {Sepiol},\ and\ \citenamefont {Lucas}}]{ballance2016high}%
  \BibitemOpen
  \bibfield  {author} {\bibinfo {author} {\bibfnamefont {C.~J.}\ \bibnamefont
  {Ballance}}, \bibinfo {author} {\bibfnamefont {T.~P.}\ \bibnamefont {Harty}},
  \bibinfo {author} {\bibfnamefont {N.~M.}\ \bibnamefont {Linke}}, \bibinfo
  {author} {\bibfnamefont {M.~A.}\ \bibnamefont {Sepiol}},\ and\ \bibinfo
  {author} {\bibfnamefont {D.~M.}\ \bibnamefont {Lucas}},\ }\bibfield  {title}
  {\bibinfo {title} {High-fidelity quantum logic gates using trapped-ion
  hyperfine qubits},\ }\href@noop {} {\bibfield  {journal} {\bibinfo  {journal}
  {Physical Review Letters}\ }\textbf {\bibinfo {volume} {117}},\ \bibinfo
  {pages} {060504} (\bibinfo {year} {2016})}\BibitemShut {NoStop}%
\bibitem [{\citenamefont {Gaebler}\ \emph {et~al.}(2016)\citenamefont
  {Gaebler}, \citenamefont {Tan}, \citenamefont {Lin}, \citenamefont {Wan},
  \citenamefont {Bowler}, \citenamefont {Keith}, \citenamefont {Glancy},
  \citenamefont {Coakley}, \citenamefont {Knill}, \citenamefont {Leibfried}
  \emph {et~al.}}]{gaebler2016high}%
  \BibitemOpen
  \bibfield  {author} {\bibinfo {author} {\bibfnamefont {J.~P.}\ \bibnamefont
  {Gaebler}}, \bibinfo {author} {\bibfnamefont {T.~R.}\ \bibnamefont {Tan}},
  \bibinfo {author} {\bibfnamefont {Y.}~\bibnamefont {Lin}}, \bibinfo {author}
  {\bibfnamefont {Y.}~\bibnamefont {Wan}}, \bibinfo {author} {\bibfnamefont
  {R.}~\bibnamefont {Bowler}}, \bibinfo {author} {\bibfnamefont {A.~C.}\
  \bibnamefont {Keith}}, \bibinfo {author} {\bibfnamefont {S.}~\bibnamefont
  {Glancy}}, \bibinfo {author} {\bibfnamefont {K.}~\bibnamefont {Coakley}},
  \bibinfo {author} {\bibfnamefont {E.}~\bibnamefont {Knill}}, \bibinfo
  {author} {\bibfnamefont {D.}~\bibnamefont {Leibfried}}, \emph {et~al.},\
  }\bibfield  {title} {\bibinfo {title} {High-fidelity universal gate set for
  {$^9$Be$^+$ } ion qubits},\ }\href@noop {} {\bibfield  {journal} {\bibinfo
  {journal} {Physical Review Letters}\ }\textbf {\bibinfo {volume} {117}},\
  \bibinfo {pages} {060505} (\bibinfo {year} {2016})}\BibitemShut {NoStop}%
\bibitem [{\citenamefont {Cai}\ \emph {et~al.}(2023)\citenamefont {Cai},
  \citenamefont {Luan}, \citenamefont {Ou}, \citenamefont {Tu}, \citenamefont
  {Yin}, \citenamefont {Zhang},\ and\ \citenamefont {Kim}}]{cai2023entangling}%
  \BibitemOpen
  \bibfield  {author} {\bibinfo {author} {\bibfnamefont {Z.}~\bibnamefont
  {Cai}}, \bibinfo {author} {\bibfnamefont {C.-Y.}\ \bibnamefont {Luan}},
  \bibinfo {author} {\bibfnamefont {L.}~\bibnamefont {Ou}}, \bibinfo {author}
  {\bibfnamefont {H.}~\bibnamefont {Tu}}, \bibinfo {author} {\bibfnamefont
  {Z.}~\bibnamefont {Yin}}, \bibinfo {author} {\bibfnamefont {J.-N.}\
  \bibnamefont {Zhang}},\ and\ \bibinfo {author} {\bibfnamefont
  {K.}~\bibnamefont {Kim}},\ }\bibfield  {title} {\bibinfo {title} {Entangling
  gates for trapped-ion quantum computation and quantum simulation},\
  }\href@noop {} {\bibfield  {journal} {\bibinfo  {journal} {Journal of the
  Korean Physical Society}\ }\textbf {\bibinfo {volume} {82}},\ \bibinfo
  {pages} {882} (\bibinfo {year} {2023})}\BibitemShut {NoStop}%
\bibitem [{\citenamefont {L{\"o}schnauer}\ \emph {et~al.}(2024)\citenamefont
  {L{\"o}schnauer}, \citenamefont {Toba}, \citenamefont {Hughes}, \citenamefont
  {King}, \citenamefont {Weber}, \citenamefont {Srinivas}, \citenamefont
  {Matt}, \citenamefont {Nourshargh}, \citenamefont {Allcock}, \citenamefont
  {Ballance} \emph {et~al.}}]{loschnauer2024scalable}%
  \BibitemOpen
  \bibfield  {author} {\bibinfo {author} {\bibfnamefont {C.}~\bibnamefont
  {L{\"o}schnauer}}, \bibinfo {author} {\bibfnamefont {J.~M.}\ \bibnamefont
  {Toba}}, \bibinfo {author} {\bibfnamefont {A.}~\bibnamefont {Hughes}},
  \bibinfo {author} {\bibfnamefont {S.}~\bibnamefont {King}}, \bibinfo {author}
  {\bibfnamefont {M.}~\bibnamefont {Weber}}, \bibinfo {author} {\bibfnamefont
  {R.}~\bibnamefont {Srinivas}}, \bibinfo {author} {\bibfnamefont
  {R.}~\bibnamefont {Matt}}, \bibinfo {author} {\bibfnamefont {R.}~\bibnamefont
  {Nourshargh}}, \bibinfo {author} {\bibfnamefont {D.}~\bibnamefont {Allcock}},
  \bibinfo {author} {\bibfnamefont {C.}~\bibnamefont {Ballance}}, \emph
  {et~al.},\ }\bibfield  {title} {\bibinfo {title} {Scalable, high-fidelity
  all-electronic control of trapped-ion qubits},\ }\href@noop {} {\bibfield
  {journal} {\bibinfo  {journal} {arXiv preprint arXiv:2407.07694}\ } (\bibinfo
  {year} {2024})}\BibitemShut {NoStop}%
\bibitem [{\citenamefont {Smith}\ \emph {et~al.}(2025)\citenamefont {Smith},
  \citenamefont {Leu}, \citenamefont {Miyanishi}, \citenamefont {Gely},\ and\
  \citenamefont {Lucas}}]{smith2025PRL}%
  \BibitemOpen
  \bibfield  {author} {\bibinfo {author} {\bibfnamefont {M.~C.}\ \bibnamefont
  {Smith}}, \bibinfo {author} {\bibfnamefont {A.~D.}\ \bibnamefont {Leu}},
  \bibinfo {author} {\bibfnamefont {K.}~\bibnamefont {Miyanishi}}, \bibinfo
  {author} {\bibfnamefont {M.~F.}\ \bibnamefont {Gely}},\ and\ \bibinfo
  {author} {\bibfnamefont {D.~M.}\ \bibnamefont {Lucas}},\ }\bibfield  {title}
  {\bibinfo {title} {Single-{{Qubit Gates}} with {{Errors}} at the
  \$\{10\}{\textasciicircum}\{{\textbackslash}ensuremath\{-\}7\}\$ {{Level}}},\
  }\href {https://doi.org/10.1103/42w2-6ccy} {\bibfield  {journal} {\bibinfo
  {journal} {Physical Review Letters}\ }\textbf {\bibinfo {volume} {134}},\
  \bibinfo {pages} {230601} (\bibinfo {year} {2025})}\BibitemShut {NoStop}%
\bibitem [{\citenamefont {Pino}\ \emph {et~al.}(2021)\citenamefont {Pino},
  \citenamefont {Dreiling}, \citenamefont {Figgatt}, \citenamefont {Gaebler},
  \citenamefont {Moses}, \citenamefont {Allman}, \citenamefont {Baldwin},
  \citenamefont {Foss-Feig}, \citenamefont {Hayes}, \citenamefont {Mayer} \emph
  {et~al.}}]{pino2021demonstration}%
  \BibitemOpen
  \bibfield  {author} {\bibinfo {author} {\bibfnamefont {J.~M.}\ \bibnamefont
  {Pino}}, \bibinfo {author} {\bibfnamefont {J.~M.}\ \bibnamefont {Dreiling}},
  \bibinfo {author} {\bibfnamefont {C.}~\bibnamefont {Figgatt}}, \bibinfo
  {author} {\bibfnamefont {J.~P.}\ \bibnamefont {Gaebler}}, \bibinfo {author}
  {\bibfnamefont {S.~A.}\ \bibnamefont {Moses}}, \bibinfo {author}
  {\bibfnamefont {M.}~\bibnamefont {Allman}}, \bibinfo {author} {\bibfnamefont
  {C.}~\bibnamefont {Baldwin}}, \bibinfo {author} {\bibfnamefont
  {M.}~\bibnamefont {Foss-Feig}}, \bibinfo {author} {\bibfnamefont
  {D.}~\bibnamefont {Hayes}}, \bibinfo {author} {\bibfnamefont
  {K.}~\bibnamefont {Mayer}}, \emph {et~al.},\ }\bibfield  {title} {\bibinfo
  {title} {Demonstration of the trapped-ion quantum ccd computer
  architecture},\ }\href@noop {} {\bibfield  {journal} {\bibinfo  {journal}
  {Nature}\ }\textbf {\bibinfo {volume} {592}},\ \bibinfo {pages} {209}
  (\bibinfo {year} {2021})}\BibitemShut {NoStop}%
\bibitem [{\citenamefont {Feng}\ \emph {et~al.}(2023)\citenamefont {Feng},
  \citenamefont {Katz}, \citenamefont {Haack}, \citenamefont {Maghrebi},
  \citenamefont {Gorshkov}, \citenamefont {Gong}, \citenamefont {Cetina},\ and\
  \citenamefont {Monroe}}]{feng2023continuous}%
  \BibitemOpen
  \bibfield  {author} {\bibinfo {author} {\bibfnamefont {L.}~\bibnamefont
  {Feng}}, \bibinfo {author} {\bibfnamefont {O.}~\bibnamefont {Katz}}, \bibinfo
  {author} {\bibfnamefont {C.}~\bibnamefont {Haack}}, \bibinfo {author}
  {\bibfnamefont {M.}~\bibnamefont {Maghrebi}}, \bibinfo {author}
  {\bibfnamefont {A.~V.}\ \bibnamefont {Gorshkov}}, \bibinfo {author}
  {\bibfnamefont {Z.}~\bibnamefont {Gong}}, \bibinfo {author} {\bibfnamefont
  {M.}~\bibnamefont {Cetina}},\ and\ \bibinfo {author} {\bibfnamefont
  {C.}~\bibnamefont {Monroe}},\ }\bibfield  {title} {\bibinfo {title}
  {Continuous symmetry breaking in a trapped-ion spin chain},\ }\href@noop {}
  {\bibfield  {journal} {\bibinfo  {journal} {Nature}\ }\textbf {\bibinfo
  {volume} {623}},\ \bibinfo {pages} {713} (\bibinfo {year}
  {2023})}\BibitemShut {NoStop}%
\bibitem [{\citenamefont {Guo}\ \emph {et~al.}(2024)\citenamefont {Guo},
  \citenamefont {Wu}, \citenamefont {Ye}, \citenamefont {Zhang}, \citenamefont
  {Lian}, \citenamefont {Yao}, \citenamefont {Wang}, \citenamefont {Yan},
  \citenamefont {Yi}, \citenamefont {Xu}, \citenamefont {Li}, \citenamefont
  {Hou}, \citenamefont {Xu}, \citenamefont {Guo}, \citenamefont {Zhang},
  \citenamefont {Qi}, \citenamefont {Zhou}, \citenamefont {He},\ and\
  \citenamefont {Duan}}]{GuoDuan2024N}%
  \BibitemOpen
  \bibfield  {author} {\bibinfo {author} {\bibfnamefont {S.-A.}\ \bibnamefont
  {Guo}}, \bibinfo {author} {\bibfnamefont {Y.-K.}\ \bibnamefont {Wu}},
  \bibinfo {author} {\bibfnamefont {J.}~\bibnamefont {Ye}}, \bibinfo {author}
  {\bibfnamefont {L.}~\bibnamefont {Zhang}}, \bibinfo {author} {\bibfnamefont
  {W.-Q.}\ \bibnamefont {Lian}}, \bibinfo {author} {\bibfnamefont
  {R.}~\bibnamefont {Yao}}, \bibinfo {author} {\bibfnamefont {Y.}~\bibnamefont
  {Wang}}, \bibinfo {author} {\bibfnamefont {R.-Y.}\ \bibnamefont {Yan}},
  \bibinfo {author} {\bibfnamefont {Y.-J.}\ \bibnamefont {Yi}}, \bibinfo
  {author} {\bibfnamefont {Y.-L.}\ \bibnamefont {Xu}}, \bibinfo {author}
  {\bibfnamefont {B.-W.}\ \bibnamefont {Li}}, \bibinfo {author} {\bibfnamefont
  {Y.-H.}\ \bibnamefont {Hou}}, \bibinfo {author} {\bibfnamefont {Y.-Z.}\
  \bibnamefont {Xu}}, \bibinfo {author} {\bibfnamefont {W.-X.}\ \bibnamefont
  {Guo}}, \bibinfo {author} {\bibfnamefont {C.}~\bibnamefont {Zhang}}, \bibinfo
  {author} {\bibfnamefont {B.-X.}\ \bibnamefont {Qi}}, \bibinfo {author}
  {\bibfnamefont {Z.-C.}\ \bibnamefont {Zhou}}, \bibinfo {author}
  {\bibfnamefont {L.}~\bibnamefont {He}},\ and\ \bibinfo {author}
  {\bibfnamefont {L.-M.}\ \bibnamefont {Duan}},\ }\bibfield  {title} {\bibinfo
  {title} {A site-resolved two-dimensional quantum simulator with hundreds of
  trapped ions},\ }\href {https://doi.org/10.1038/s41586-024-07459-0}
  {\bibfield  {journal} {\bibinfo  {journal} {Nature}\ }\textbf {\bibinfo
  {volume} {630}},\ \bibinfo {pages} {613} (\bibinfo {year}
  {2024})}\BibitemShut {NoStop}%
\bibitem [{\citenamefont {Kiesenhofer}\ \emph {et~al.}(2023)\citenamefont
  {Kiesenhofer}, \citenamefont {Hainzer}, \citenamefont {Zhdanov},
  \citenamefont {Holz}, \citenamefont {Bock}, \citenamefont {Ollikainen},\ and\
  \citenamefont {Roos}}]{kiesenhofer2023controlling}%
  \BibitemOpen
  \bibfield  {author} {\bibinfo {author} {\bibfnamefont {D.}~\bibnamefont
  {Kiesenhofer}}, \bibinfo {author} {\bibfnamefont {H.}~\bibnamefont
  {Hainzer}}, \bibinfo {author} {\bibfnamefont {A.}~\bibnamefont {Zhdanov}},
  \bibinfo {author} {\bibfnamefont {P.~C.}\ \bibnamefont {Holz}}, \bibinfo
  {author} {\bibfnamefont {M.}~\bibnamefont {Bock}}, \bibinfo {author}
  {\bibfnamefont {T.}~\bibnamefont {Ollikainen}},\ and\ \bibinfo {author}
  {\bibfnamefont {C.~F.}\ \bibnamefont {Roos}},\ }\bibfield  {title} {\bibinfo
  {title} {Controlling two-dimensional coulomb crystals of more than 100 ions
  in a monolithic radio-frequency trap},\ }\href@noop {} {\bibfield  {journal}
  {\bibinfo  {journal} {PRX Quantum}\ }\textbf {\bibinfo {volume} {4}},\
  \bibinfo {pages} {020317} (\bibinfo {year} {2023})}\BibitemShut {NoStop}%
\bibitem [{\citenamefont {Shankar}\ \emph {et~al.}(2022)\citenamefont
  {Shankar}, \citenamefont {Yuzbashyan}, \citenamefont {Gurarie}, \citenamefont
  {Zoller}, \citenamefont {Bollinger},\ and\ \citenamefont
  {Rey}}]{shankar2022simulating}%
  \BibitemOpen
  \bibfield  {author} {\bibinfo {author} {\bibfnamefont {A.}~\bibnamefont
  {Shankar}}, \bibinfo {author} {\bibfnamefont {E.~A.}\ \bibnamefont
  {Yuzbashyan}}, \bibinfo {author} {\bibfnamefont {V.}~\bibnamefont {Gurarie}},
  \bibinfo {author} {\bibfnamefont {P.}~\bibnamefont {Zoller}}, \bibinfo
  {author} {\bibfnamefont {J.~J.}\ \bibnamefont {Bollinger}},\ and\ \bibinfo
  {author} {\bibfnamefont {A.~M.}\ \bibnamefont {Rey}},\ }\bibfield  {title}
  {\bibinfo {title} {Simulating dynamical phases of chiral p+ ip
  superconductors with a trapped ion magnet},\ }\href@noop {} {\bibfield
  {journal} {\bibinfo  {journal} {PRX Quantum}\ }\textbf {\bibinfo {volume}
  {3}},\ \bibinfo {pages} {040324} (\bibinfo {year} {2022})}\BibitemShut
  {NoStop}%
\bibitem [{\citenamefont {Lu}\ \emph {et~al.}(2019)\citenamefont {Lu},
  \citenamefont {Zhang}, \citenamefont {Zhang}, \citenamefont {Chen},
  \citenamefont {Shen}, \citenamefont {Zhang}, \citenamefont {Zhang},\ and\
  \citenamefont {Kim}}]{LuKim2019N}%
  \BibitemOpen
  \bibfield  {author} {\bibinfo {author} {\bibfnamefont {Y.}~\bibnamefont
  {Lu}}, \bibinfo {author} {\bibfnamefont {S.}~\bibnamefont {Zhang}}, \bibinfo
  {author} {\bibfnamefont {K.}~\bibnamefont {Zhang}}, \bibinfo {author}
  {\bibfnamefont {W.}~\bibnamefont {Chen}}, \bibinfo {author} {\bibfnamefont
  {Y.}~\bibnamefont {Shen}}, \bibinfo {author} {\bibfnamefont {J.}~\bibnamefont
  {Zhang}}, \bibinfo {author} {\bibfnamefont {J.-N.}\ \bibnamefont {Zhang}},\
  and\ \bibinfo {author} {\bibfnamefont {K.}~\bibnamefont {Kim}},\ }\bibfield
  {title} {\bibinfo {title} {Global entangling gates on arbitrary ion qubits},\
  }\href {https://doi.org/10.1038/s41586-019-1428-4} {\bibfield  {journal}
  {\bibinfo  {journal} {Nature}\ }\textbf {\bibinfo {volume} {572}},\ \bibinfo
  {pages} {363} (\bibinfo {year} {2019})}\BibitemShut {NoStop}%
\bibitem [{\citenamefont {Figgatt}\ \emph {et~al.}(2019)\citenamefont
  {Figgatt}, \citenamefont {Ostrander}, \citenamefont {Linke}, \citenamefont
  {Landsman}, \citenamefont {Zhu}, \citenamefont {Maslov},\ and\ \citenamefont
  {Monroe}}]{FiggattMonroe2019N}%
  \BibitemOpen
  \bibfield  {author} {\bibinfo {author} {\bibfnamefont {C.}~\bibnamefont
  {Figgatt}}, \bibinfo {author} {\bibfnamefont {A.}~\bibnamefont {Ostrander}},
  \bibinfo {author} {\bibfnamefont {N.~M.}\ \bibnamefont {Linke}}, \bibinfo
  {author} {\bibfnamefont {K.~A.}\ \bibnamefont {Landsman}}, \bibinfo {author}
  {\bibfnamefont {D.}~\bibnamefont {Zhu}}, \bibinfo {author} {\bibfnamefont
  {D.}~\bibnamefont {Maslov}},\ and\ \bibinfo {author} {\bibfnamefont
  {C.}~\bibnamefont {Monroe}},\ }\bibfield  {title} {\bibinfo {title} {Parallel
  entangling operations on a universal ion-trap quantum computer},\ }\href
  {https://doi.org/10.1038/s41586-019-1427-5} {\bibfield  {journal} {\bibinfo
  {journal} {Nature}\ ,\ \bibinfo {pages} {1}} (\bibinfo {year}
  {2019})}\BibitemShut {NoStop}%
\bibitem [{\citenamefont {Landsman}\ \emph {et~al.}(2019)\citenamefont
  {Landsman}, \citenamefont {Wu}, \citenamefont {Leung}, \citenamefont {Zhu},
  \citenamefont {Linke}, \citenamefont {Brown}, \citenamefont {Duan},\ and\
  \citenamefont {Monroe}}]{landsman2019two}%
  \BibitemOpen
  \bibfield  {author} {\bibinfo {author} {\bibfnamefont {K.~A.}\ \bibnamefont
  {Landsman}}, \bibinfo {author} {\bibfnamefont {Y.}~\bibnamefont {Wu}},
  \bibinfo {author} {\bibfnamefont {P.~H.}\ \bibnamefont {Leung}}, \bibinfo
  {author} {\bibfnamefont {D.}~\bibnamefont {Zhu}}, \bibinfo {author}
  {\bibfnamefont {N.~M.}\ \bibnamefont {Linke}}, \bibinfo {author}
  {\bibfnamefont {K.~R.}\ \bibnamefont {Brown}}, \bibinfo {author}
  {\bibfnamefont {L.}~\bibnamefont {Duan}},\ and\ \bibinfo {author}
  {\bibfnamefont {C.}~\bibnamefont {Monroe}},\ }\bibfield  {title} {\bibinfo
  {title} {Two-qubit entangling gates within arbitrarily long chains of trapped
  ions},\ }\href@noop {} {\bibfield  {journal} {\bibinfo  {journal} {Physical
  Review A}\ }\textbf {\bibinfo {volume} {100}},\ \bibinfo {pages} {022332}
  (\bibinfo {year} {2019})}\BibitemShut {NoStop}%
\bibitem [{\citenamefont {Chen}\ \emph
  {et~al.}(2024{\natexlab{a}})\citenamefont {Chen}, \citenamefont {Nielsen},
  \citenamefont {Ebert}, \citenamefont {Inlek}, \citenamefont {Wright},
  \citenamefont {Chaplin}, \citenamefont {Maksymov}, \citenamefont {P{\'a}ez},
  \citenamefont {Poudel}, \citenamefont {Maunz} \emph
  {et~al.}}]{chen2024benchmarking}%
  \BibitemOpen
  \bibfield  {author} {\bibinfo {author} {\bibfnamefont {J.-S.}\ \bibnamefont
  {Chen}}, \bibinfo {author} {\bibfnamefont {E.}~\bibnamefont {Nielsen}},
  \bibinfo {author} {\bibfnamefont {M.}~\bibnamefont {Ebert}}, \bibinfo
  {author} {\bibfnamefont {V.}~\bibnamefont {Inlek}}, \bibinfo {author}
  {\bibfnamefont {K.}~\bibnamefont {Wright}}, \bibinfo {author} {\bibfnamefont
  {V.}~\bibnamefont {Chaplin}}, \bibinfo {author} {\bibfnamefont
  {A.}~\bibnamefont {Maksymov}}, \bibinfo {author} {\bibfnamefont
  {E.}~\bibnamefont {P{\'a}ez}}, \bibinfo {author} {\bibfnamefont
  {A.}~\bibnamefont {Poudel}}, \bibinfo {author} {\bibfnamefont
  {P.}~\bibnamefont {Maunz}}, \emph {et~al.},\ }\bibfield  {title} {\bibinfo
  {title} {Benchmarking a trapped-ion quantum computer with 30 qubits},\
  }\href@noop {} {\bibfield  {journal} {\bibinfo  {journal} {Quantum}\ }\textbf
  {\bibinfo {volume} {8}},\ \bibinfo {pages} {1516} (\bibinfo {year}
  {2024}{\natexlab{a}})}\BibitemShut {NoStop}%
\bibitem [{\citenamefont {Hou}\ \emph {et~al.}(2024)\citenamefont {Hou},
  \citenamefont {Yi}, \citenamefont {Wu}, \citenamefont {Chen}, \citenamefont
  {Zhang}, \citenamefont {Wang}, \citenamefont {Xu}, \citenamefont {Zhang},
  \citenamefont {Mei}, \citenamefont {Yang}, \citenamefont {Ma}, \citenamefont
  {Guo}, \citenamefont {Ye}, \citenamefont {Qi}, \citenamefont {Zhou},
  \citenamefont {Hou},\ and\ \citenamefont {Duan}}]{HouDuan2024NC}%
  \BibitemOpen
  \bibfield  {author} {\bibinfo {author} {\bibfnamefont {Y.-H.}\ \bibnamefont
  {Hou}}, \bibinfo {author} {\bibfnamefont {Y.-J.}\ \bibnamefont {Yi}},
  \bibinfo {author} {\bibfnamefont {Y.-K.}\ \bibnamefont {Wu}}, \bibinfo
  {author} {\bibfnamefont {Y.-Y.}\ \bibnamefont {Chen}}, \bibinfo {author}
  {\bibfnamefont {L.}~\bibnamefont {Zhang}}, \bibinfo {author} {\bibfnamefont
  {Y.}~\bibnamefont {Wang}}, \bibinfo {author} {\bibfnamefont {Y.-L.}\
  \bibnamefont {Xu}}, \bibinfo {author} {\bibfnamefont {C.}~\bibnamefont
  {Zhang}}, \bibinfo {author} {\bibfnamefont {Q.-X.}\ \bibnamefont {Mei}},
  \bibinfo {author} {\bibfnamefont {H.-X.}\ \bibnamefont {Yang}}, \bibinfo
  {author} {\bibfnamefont {J.-Y.}\ \bibnamefont {Ma}}, \bibinfo {author}
  {\bibfnamefont {S.-A.}\ \bibnamefont {Guo}}, \bibinfo {author} {\bibfnamefont
  {J.}~\bibnamefont {Ye}}, \bibinfo {author} {\bibfnamefont {B.-X.}\
  \bibnamefont {Qi}}, \bibinfo {author} {\bibfnamefont {Z.-C.}\ \bibnamefont
  {Zhou}}, \bibinfo {author} {\bibfnamefont {P.-Y.}\ \bibnamefont {Hou}},\ and\
  \bibinfo {author} {\bibfnamefont {L.-M.}\ \bibnamefont {Duan}},\ }\bibfield
  {title} {\bibinfo {title} {Individually addressed entangling gates in a
  two-dimensional ion crystal},\ }\href
  {https://doi.org/10.1038/s41467-024-53405-z} {\bibfield  {journal} {\bibinfo
  {journal} {Nature Communications}\ }\textbf {\bibinfo {volume} {15}},\
  \bibinfo {pages} {9710} (\bibinfo {year} {2024})}\BibitemShut {NoStop}%
\bibitem [{\citenamefont {Olsacher}\ \emph {et~al.}(2020)\citenamefont
  {Olsacher}, \citenamefont {Postler}, \citenamefont {Schindler}, \citenamefont
  {Monz}, \citenamefont {Zoller},\ and\ \citenamefont
  {Sieberer}}]{olsacher2020scalable}%
  \BibitemOpen
  \bibfield  {author} {\bibinfo {author} {\bibfnamefont {T.}~\bibnamefont
  {Olsacher}}, \bibinfo {author} {\bibfnamefont {L.}~\bibnamefont {Postler}},
  \bibinfo {author} {\bibfnamefont {P.}~\bibnamefont {Schindler}}, \bibinfo
  {author} {\bibfnamefont {T.}~\bibnamefont {Monz}}, \bibinfo {author}
  {\bibfnamefont {P.}~\bibnamefont {Zoller}},\ and\ \bibinfo {author}
  {\bibfnamefont {L.~M.}\ \bibnamefont {Sieberer}},\ }\bibfield  {title}
  {\bibinfo {title} {Scalable and parallel tweezer gates for quantum computing
  with long ion strings},\ }\href@noop {} {\bibfield  {journal} {\bibinfo
  {journal} {PRX Quantum}\ }\textbf {\bibinfo {volume} {1}},\ \bibinfo {pages}
  {020316} (\bibinfo {year} {2020})}\BibitemShut {NoStop}%
\bibitem [{\citenamefont {Shen}\ and\ \citenamefont {Lin}(2020)}]{shen2020NJP}%
  \BibitemOpen
  \bibfield  {author} {\bibinfo {author} {\bibfnamefont {Y.-C.}\ \bibnamefont
  {Shen}}\ and\ \bibinfo {author} {\bibfnamefont {G.-D.}\ \bibnamefont {Lin}},\
  }\bibfield  {title} {\bibinfo {title} {Scalable quantum computing stabilised
  by optical tweezers on an ion crystal},\ }\href
  {https://dx.doi.org/10.1088/1367-2630/ab84b6} {\bibfield  {journal} {\bibinfo
   {journal} {New Journal of Physics}\ }\textbf {\bibinfo {volume} {22}},\
  \bibinfo {pages} {053032} (\bibinfo {year} {2020})}\BibitemShut {NoStop}%
\bibitem [{\citenamefont {Mazzanti}\ \emph {et~al.}(2021)\citenamefont
  {Mazzanti}, \citenamefont {Sch{\"u}ssler}, \citenamefont {Arias~Espinoza},
  \citenamefont {Wu}, \citenamefont {Gerritsma},\ and\ \citenamefont
  {{Safavi-Naini}}}]{mazzanti2021PRL}%
  \BibitemOpen
  \bibfield  {author} {\bibinfo {author} {\bibfnamefont {M.}~\bibnamefont
  {Mazzanti}}, \bibinfo {author} {\bibfnamefont {R.~X.}\ \bibnamefont
  {Sch{\"u}ssler}}, \bibinfo {author} {\bibfnamefont {J.~D.}\ \bibnamefont
  {Arias~Espinoza}}, \bibinfo {author} {\bibfnamefont {Z.}~\bibnamefont {Wu}},
  \bibinfo {author} {\bibfnamefont {R.}~\bibnamefont {Gerritsma}},\ and\
  \bibinfo {author} {\bibfnamefont {A.}~\bibnamefont {{Safavi-Naini}}},\
  }\bibfield  {title} {\bibinfo {title} {Trapped {{Ion Quantum Computing Using
  Optical Tweezers}} and {{Electric Fields}}},\ }\href
  {https://doi.org/10.1103/PhysRevLett.127.260502} {\bibfield  {journal}
  {\bibinfo  {journal} {Physical Review Letters}\ }\textbf {\bibinfo {volume}
  {127}},\ \bibinfo {pages} {260502} (\bibinfo {year} {2021})}\BibitemShut
  {NoStop}%
\bibitem [{\citenamefont {Teoh}\ \emph {et~al.}(2021)\citenamefont {Teoh},
  \citenamefont {Sajjan}, \citenamefont {Sun}, \citenamefont {Rajabi},\ and\
  \citenamefont {Islam}}]{teoh2021manipulating}%
  \BibitemOpen
  \bibfield  {author} {\bibinfo {author} {\bibfnamefont {Y.~H.}\ \bibnamefont
  {Teoh}}, \bibinfo {author} {\bibfnamefont {M.}~\bibnamefont {Sajjan}},
  \bibinfo {author} {\bibfnamefont {Z.}~\bibnamefont {Sun}}, \bibinfo {author}
  {\bibfnamefont {F.}~\bibnamefont {Rajabi}},\ and\ \bibinfo {author}
  {\bibfnamefont {R.}~\bibnamefont {Islam}},\ }\bibfield  {title} {\bibinfo
  {title} {Manipulating phonons of a trapped-ion system using optical
  tweezers},\ }\href@noop {} {\bibfield  {journal} {\bibinfo  {journal}
  {Physical Review A}\ }\textbf {\bibinfo {volume} {104}},\ \bibinfo {pages}
  {022420} (\bibinfo {year} {2021})}\BibitemShut {NoStop}%
\bibitem [{\citenamefont {Schwerdt}\ \emph {et~al.}(2024)\citenamefont
  {Schwerdt}, \citenamefont {Peleg}, \citenamefont {Shapira}, \citenamefont
  {Priel}, \citenamefont {Florshaim}, \citenamefont {Gross}, \citenamefont
  {Zalic}, \citenamefont {Afek}, \citenamefont {Akerman}, \citenamefont {Stern}
  \emph {et~al.}}]{schwerdt2024scalable}%
  \BibitemOpen
  \bibfield  {author} {\bibinfo {author} {\bibfnamefont {D.}~\bibnamefont
  {Schwerdt}}, \bibinfo {author} {\bibfnamefont {L.}~\bibnamefont {Peleg}},
  \bibinfo {author} {\bibfnamefont {Y.}~\bibnamefont {Shapira}}, \bibinfo
  {author} {\bibfnamefont {N.}~\bibnamefont {Priel}}, \bibinfo {author}
  {\bibfnamefont {Y.}~\bibnamefont {Florshaim}}, \bibinfo {author}
  {\bibfnamefont {A.}~\bibnamefont {Gross}}, \bibinfo {author} {\bibfnamefont
  {A.}~\bibnamefont {Zalic}}, \bibinfo {author} {\bibfnamefont
  {G.}~\bibnamefont {Afek}}, \bibinfo {author} {\bibfnamefont {N.}~\bibnamefont
  {Akerman}}, \bibinfo {author} {\bibfnamefont {A.}~\bibnamefont {Stern}},
  \emph {et~al.},\ }\bibfield  {title} {\bibinfo {title} {Scalable architecture
  for trapped-ion quantum computing using rf traps and dynamic optical
  potentials},\ }\href@noop {} {\bibfield  {journal} {\bibinfo  {journal}
  {Physical Review X}\ }\textbf {\bibinfo {volume} {14}},\ \bibinfo {pages}
  {041017} (\bibinfo {year} {2024})}\BibitemShut {NoStop}%
\bibitem [{\citenamefont {Cheng}\ \emph {et~al.}(2024)\citenamefont {Cheng},
  \citenamefont {Liu}, \citenamefont {Peng},\ and\ \citenamefont
  {Gong}}]{ChengGong2024PRA}%
  \BibitemOpen
  \bibfield  {author} {\bibinfo {author} {\bibfnamefont {L.}~\bibnamefont
  {Cheng}}, \bibinfo {author} {\bibfnamefont {S.-C.}\ \bibnamefont {Liu}},
  \bibinfo {author} {\bibfnamefont {L.-Y.}\ \bibnamefont {Peng}},\ and\
  \bibinfo {author} {\bibfnamefont {Q.}~\bibnamefont {Gong}},\ }\bibfield
  {title} {\bibinfo {title} {Crosstalk suppression of parallel gates for
  fault-tolerant quantum computation with trapped ions via optical tweezers},\
  }\href {https://doi.org/10.1103/PhysRevApplied.22.034021} {\bibfield
  {journal} {\bibinfo  {journal} {Physical Review Applied}\ }\textbf {\bibinfo
  {volume} {22}},\ \bibinfo {pages} {034021} (\bibinfo {year}
  {2024})}\BibitemShut {NoStop}%
\bibitem [{\citenamefont {Spreeuw}(2020)}]{SpreeuwSpreeuw2020PRL}%
  \BibitemOpen
  \bibfield  {author} {\bibinfo {author} {\bibfnamefont {R.~J.~C.}\
  \bibnamefont {Spreeuw}},\ }\bibfield  {title} {\bibinfo {title} {Off-{{Axis
  Dipole Forces}} in {{Optical Tweezers}} by an {{Optical Analog}} of the
  {{Magnus Effect}}},\ }\href {https://doi.org/10.1103/PhysRevLett.125.233201}
  {\bibfield  {journal} {\bibinfo  {journal} {Physical Review Letters}\
  }\textbf {\bibinfo {volume} {125}},\ \bibinfo {pages} {233201} (\bibinfo
  {year} {2020})}\BibitemShut {NoStop}%
\bibitem [{\citenamefont {Mazzanti}\ \emph {et~al.}(2023)\citenamefont
  {Mazzanti}, \citenamefont {Gerritsma}, \citenamefont {Spreeuw},\ and\
  \citenamefont {{Safavi-Naini}}}]{MazzantiSafaviNaini2023PRR}%
  \BibitemOpen
  \bibfield  {author} {\bibinfo {author} {\bibfnamefont {M.}~\bibnamefont
  {Mazzanti}}, \bibinfo {author} {\bibfnamefont {R.}~\bibnamefont {Gerritsma}},
  \bibinfo {author} {\bibfnamefont {R.~J.~C.}\ \bibnamefont {Spreeuw}},\ and\
  \bibinfo {author} {\bibfnamefont {A.}~\bibnamefont {{Safavi-Naini}}},\
  }\bibfield  {title} {\bibinfo {title} {Trapped ions quantum logic gate with
  optical tweezers and the {{Magnus}} effect},\ }\href
  {https://doi.org/10.1103/PhysRevResearch.5.033036} {\bibfield  {journal}
  {\bibinfo  {journal} {Physical Review Research}\ }\textbf {\bibinfo {volume}
  {5}},\ \bibinfo {pages} {033036} (\bibinfo {year} {2023})}\BibitemShut
  {NoStop}%
\bibitem [{\citenamefont {Leibfried}\ \emph {et~al.}(2003)\citenamefont
  {Leibfried}, \citenamefont {Blatt}, \citenamefont {Monroe},\ and\
  \citenamefont {Wineland}}]{LeibfriedWineland2003RoMP}%
  \BibitemOpen
  \bibfield  {author} {\bibinfo {author} {\bibfnamefont {D.}~\bibnamefont
  {Leibfried}}, \bibinfo {author} {\bibfnamefont {R.}~\bibnamefont {Blatt}},
  \bibinfo {author} {\bibfnamefont {C.}~\bibnamefont {Monroe}},\ and\ \bibinfo
  {author} {\bibfnamefont {D.}~\bibnamefont {Wineland}},\ }\bibfield  {title}
  {\bibinfo {title} {Quantum dynamics of single trapped ions},\ }\href
  {https://doi.org/10.1103/RevModPhys.75.281} {\bibfield  {journal} {\bibinfo
  {journal} {Reviews of Modern Physics}\ }\textbf {\bibinfo {volume} {75}},\
  \bibinfo {pages} {281} (\bibinfo {year} {2003})}\BibitemShut {NoStop}%
\bibitem [{\citenamefont {Saner}\ \emph {et~al.}(2023)\citenamefont {Saner},
  \citenamefont {B{\u a}z{\u a}van}, \citenamefont {Minder}, \citenamefont
  {Drmota}, \citenamefont {Webb}, \citenamefont {Araneda}, \citenamefont
  {Srinivas}, \citenamefont {Lucas},\ and\ \citenamefont
  {Ballance}}]{SanerBallance2023PRL}%
  \BibitemOpen
  \bibfield  {author} {\bibinfo {author} {\bibfnamefont {S.}~\bibnamefont
  {Saner}}, \bibinfo {author} {\bibfnamefont {O.}~\bibnamefont {B{\u a}z{\u
  a}van}}, \bibinfo {author} {\bibfnamefont {M.}~\bibnamefont {Minder}},
  \bibinfo {author} {\bibfnamefont {P.}~\bibnamefont {Drmota}}, \bibinfo
  {author} {\bibfnamefont {D.~J.}\ \bibnamefont {Webb}}, \bibinfo {author}
  {\bibfnamefont {G.}~\bibnamefont {Araneda}}, \bibinfo {author} {\bibfnamefont
  {R.}~\bibnamefont {Srinivas}}, \bibinfo {author} {\bibfnamefont {D.~M.}\
  \bibnamefont {Lucas}},\ and\ \bibinfo {author} {\bibfnamefont {C.~J.}\
  \bibnamefont {Ballance}},\ }\bibfield  {title} {\bibinfo {title} {Breaking
  the {{Entangling Gate Speed Limit}} for {{Trapped-Ion Qubits Using}} a
  {{Phase-Stable Standing Wave}}},\ }\href
  {https://doi.org/10.1103/PhysRevLett.131.220601} {\bibfield  {journal}
  {\bibinfo  {journal} {Physical Review Letters}\ }\textbf {\bibinfo {volume}
  {131}},\ \bibinfo {pages} {220601} (\bibinfo {year} {2023})}\BibitemShut
  {NoStop}%
\bibitem [{\citenamefont {Nape}\ \emph {et~al.}(2022)\citenamefont {Nape},
  \citenamefont {Singh}, \citenamefont {Klug}, \citenamefont {Buono},
  \citenamefont {{Rosales-Guzman}}, \citenamefont {McWilliam}, \citenamefont
  {{Franke-Arnold}}, \citenamefont {Kritzinger}, \citenamefont {Forbes},
  \citenamefont {Dudley},\ and\ \citenamefont {Forbes}}]{nape2022NP}%
  \BibitemOpen
  \bibfield  {author} {\bibinfo {author} {\bibfnamefont {I.}~\bibnamefont
  {Nape}}, \bibinfo {author} {\bibfnamefont {K.}~\bibnamefont {Singh}},
  \bibinfo {author} {\bibfnamefont {A.}~\bibnamefont {Klug}}, \bibinfo {author}
  {\bibfnamefont {W.}~\bibnamefont {Buono}}, \bibinfo {author} {\bibfnamefont
  {C.}~\bibnamefont {{Rosales-Guzman}}}, \bibinfo {author} {\bibfnamefont
  {A.}~\bibnamefont {McWilliam}}, \bibinfo {author} {\bibfnamefont
  {S.}~\bibnamefont {{Franke-Arnold}}}, \bibinfo {author} {\bibfnamefont
  {A.}~\bibnamefont {Kritzinger}}, \bibinfo {author} {\bibfnamefont
  {P.}~\bibnamefont {Forbes}}, \bibinfo {author} {\bibfnamefont
  {A.}~\bibnamefont {Dudley}},\ and\ \bibinfo {author} {\bibfnamefont
  {A.}~\bibnamefont {Forbes}},\ }\bibfield  {title} {\bibinfo {title}
  {Revealing the invariance of vectorial structured light in complex media},\
  }\href {https://doi.org/10.1038/s41566-022-01023-w} {\bibfield  {journal}
  {\bibinfo  {journal} {Nature Photonics}\ }\textbf {\bibinfo {volume} {16}},\
  \bibinfo {pages} {538} (\bibinfo {year} {2022})}\BibitemShut {NoStop}%
\bibitem [{\citenamefont {Wang}\ \emph
  {et~al.}(2020{\natexlab{a}})\citenamefont {Wang}, \citenamefont
  {Castellucci},\ and\ \citenamefont {{Franke-Arnold}}}]{wang2020AQS}%
  \BibitemOpen
  \bibfield  {author} {\bibinfo {author} {\bibfnamefont {J.}~\bibnamefont
  {Wang}}, \bibinfo {author} {\bibfnamefont {F.}~\bibnamefont {Castellucci}},\
  and\ \bibinfo {author} {\bibfnamefont {S.}~\bibnamefont {{Franke-Arnold}}},\
  }\bibfield  {title} {\bibinfo {title} {Vectorial light--matter interaction:
  {{Exploring}} spatially structured complex light fields},\ }\href
  {https://doi.org/10.1116/5.0016007} {\bibfield  {journal} {\bibinfo
  {journal} {AVS Quantum Science}\ }\textbf {\bibinfo {volume} {2}},\ \bibinfo
  {pages} {031702} (\bibinfo {year} {2020}{\natexlab{a}})}\BibitemShut
  {NoStop}%
\bibitem [{\citenamefont {He}\ \emph {et~al.}(2021)\citenamefont {He},
  \citenamefont {Cui}, \citenamefont {Li}, \citenamefont {Qian}, \citenamefont
  {Chen}, \citenamefont {Ai}, \citenamefont {Huang}, \citenamefont {Li},\ and\
  \citenamefont {Guo}}]{HeGuo2021RSI}%
  \BibitemOpen
  \bibfield  {author} {\bibinfo {author} {\bibfnamefont {R.}~\bibnamefont
  {He}}, \bibinfo {author} {\bibfnamefont {J.-M.}\ \bibnamefont {Cui}},
  \bibinfo {author} {\bibfnamefont {R.-R.}\ \bibnamefont {Li}}, \bibinfo
  {author} {\bibfnamefont {Z.-H.}\ \bibnamefont {Qian}}, \bibinfo {author}
  {\bibfnamefont {Y.}~\bibnamefont {Chen}}, \bibinfo {author} {\bibfnamefont
  {M.-Z.}\ \bibnamefont {Ai}}, \bibinfo {author} {\bibfnamefont {Y.-F.}\
  \bibnamefont {Huang}}, \bibinfo {author} {\bibfnamefont {C.-F.}\ \bibnamefont
  {Li}},\ and\ \bibinfo {author} {\bibfnamefont {G.-C.}\ \bibnamefont {Guo}},\
  }\bibfield  {title} {\bibinfo {title} {An ion trap apparatus with high
  optical access in multiple directions},\ }\href
  {https://doi.org/10.1063/5.0043985} {\bibfield  {journal} {\bibinfo
  {journal} {Rev. Sci. Instrum.}\ }\textbf {\bibinfo {volume} {92}},\ \bibinfo
  {pages} {073201} (\bibinfo {year} {2021})},\ \bibinfo {note} {publisher:
  American Institute of Physics}\BibitemShut {NoStop}%
\bibitem [{\citenamefont {Olmschenk}\ \emph {et~al.}(2007)\citenamefont
  {Olmschenk}, \citenamefont {Younge}, \citenamefont {Moehring}, \citenamefont
  {Matsukevich}, \citenamefont {Maunz},\ and\ \citenamefont
  {Monroe}}]{OlmschenkMonroe2007PRA}%
  \BibitemOpen
  \bibfield  {author} {\bibinfo {author} {\bibfnamefont {S.}~\bibnamefont
  {Olmschenk}}, \bibinfo {author} {\bibfnamefont {K.~C.}\ \bibnamefont
  {Younge}}, \bibinfo {author} {\bibfnamefont {D.~L.}\ \bibnamefont
  {Moehring}}, \bibinfo {author} {\bibfnamefont {D.~N.}\ \bibnamefont
  {Matsukevich}}, \bibinfo {author} {\bibfnamefont {P.}~\bibnamefont {Maunz}},\
  and\ \bibinfo {author} {\bibfnamefont {C.}~\bibnamefont {Monroe}},\
  }\bibfield  {title} {\bibinfo {title} {Manipulation and detection of a
  trapped {{Yb}}{\textasciicircum}\{+\} hyperfine qubit},\ }\href
  {https://doi.org/10.1103/PhysRevA.76.052314} {\bibfield  {journal} {\bibinfo
  {journal} {Physical Review A}\ }\textbf {\bibinfo {volume} {76}},\ \bibinfo
  {pages} {052314} (\bibinfo {year} {2007})}\BibitemShut {NoStop}%
\bibitem [{\citenamefont {Shen}\ and\ \citenamefont
  {Duan}(2012)}]{shen2012correcting}%
  \BibitemOpen
  \bibfield  {author} {\bibinfo {author} {\bibfnamefont {C.}~\bibnamefont
  {Shen}}\ and\ \bibinfo {author} {\bibfnamefont {L.~M.}\ \bibnamefont
  {Duan}},\ }\bibfield  {title} {\bibinfo {title} {Correcting detection errors
  in quantum state engineering through data processing},\ }\href@noop {}
  {\bibfield  {journal} {\bibinfo  {journal} {New Journal of Physics}\ }\textbf
  {\bibinfo {volume} {14}},\ \bibinfo {pages} {053053} (\bibinfo {year}
  {2012})}\BibitemShut {NoStop}%
\bibitem [{Note1()}]{Note1}%
  \BibitemOpen
  \bibinfo {note} {See Supplemental Material at {[}URL{]} for details of the
  laser system, details of ion-laser alignment}\BibitemShut {NoStop}%
\bibitem [{\citenamefont {Li}\ \emph {et~al.}(2022)\citenamefont {Li},
  \citenamefont {He}, \citenamefont {Cui}, \citenamefont {Chen}, \citenamefont
  {Ye}, \citenamefont {Chen}, \citenamefont {Huang}, \citenamefont {Li},\ and\
  \citenamefont {Guo}}]{LiGuo2022OE}%
  \BibitemOpen
  \bibfield  {author} {\bibinfo {author} {\bibfnamefont {R.-R.}\ \bibnamefont
  {Li}}, \bibinfo {author} {\bibfnamefont {R.}~\bibnamefont {He}}, \bibinfo
  {author} {\bibfnamefont {J.-M.}\ \bibnamefont {Cui}}, \bibinfo {author}
  {\bibfnamefont {Y.}~\bibnamefont {Chen}}, \bibinfo {author} {\bibfnamefont
  {W.-R.}\ \bibnamefont {Ye}}, \bibinfo {author} {\bibfnamefont {Y.-L.}\
  \bibnamefont {Chen}}, \bibinfo {author} {\bibfnamefont {Y.-F.}\ \bibnamefont
  {Huang}}, \bibinfo {author} {\bibfnamefont {C.-F.}\ \bibnamefont {Li}},\ and\
  \bibinfo {author} {\bibfnamefont {G.-C.}\ \bibnamefont {Guo}},\ }\bibfield
  {title} {\bibinfo {title} {A versatile multi-tone laser system for
  manipulating atomic qubits based on a fiber {{Mach}}--{{Zehnder}} modulator
  and second harmonic generation},\ }\href {https://doi.org/10.1364/OE.462737}
  {\bibfield  {journal} {\bibinfo  {journal} {Optics Express}\ }\textbf
  {\bibinfo {volume} {30}},\ \bibinfo {pages} {30098} (\bibinfo {year}
  {2022})}\BibitemShut {NoStop}%
\bibitem [{\citenamefont {Chen}\ \emph
  {et~al.}(2024{\natexlab{b}})\citenamefont {Chen}, \citenamefont {Li},
  \citenamefont {He}, \citenamefont {Chen}, \citenamefont {Qi}, \citenamefont
  {Cui}, \citenamefont {Huang}, \citenamefont {Li},\ and\ \citenamefont
  {Guo}}]{ChenGuo2024PRA}%
  \BibitemOpen
  \bibfield  {author} {\bibinfo {author} {\bibfnamefont {Y.-L.}\ \bibnamefont
  {Chen}}, \bibinfo {author} {\bibfnamefont {R.-R.}\ \bibnamefont {Li}},
  \bibinfo {author} {\bibfnamefont {R.}~\bibnamefont {He}}, \bibinfo {author}
  {\bibfnamefont {S.-Q.}\ \bibnamefont {Chen}}, \bibinfo {author}
  {\bibfnamefont {W.-H.}\ \bibnamefont {Qi}}, \bibinfo {author} {\bibfnamefont
  {J.-M.}\ \bibnamefont {Cui}}, \bibinfo {author} {\bibfnamefont {Y.-F.}\
  \bibnamefont {Huang}}, \bibinfo {author} {\bibfnamefont {C.-F.}\ \bibnamefont
  {Li}},\ and\ \bibinfo {author} {\bibfnamefont {G.-C.}\ \bibnamefont {Guo}},\
  }\bibfield  {title} {\bibinfo {title} {Low-crosstalk optical addressing
  system for atomic qubits based on multiple objectives and acousto-optic
  deflectors},\ }\href {https://doi.org/10.1103/PhysRevApplied.22.054003}
  {\bibfield  {journal} {\bibinfo  {journal} {Physical Review Applied}\
  }\textbf {\bibinfo {volume} {22}},\ \bibinfo {pages} {054003} (\bibinfo
  {year} {2024}{\natexlab{b}})}\BibitemShut {NoStop}%
\bibitem [{\citenamefont {Shapira}\ \emph {et~al.}(2023)\citenamefont
  {Shapira}, \citenamefont {Cohen}, \citenamefont {Akerman}, \citenamefont
  {Stern},\ and\ \citenamefont {Ozeri}}]{shapira2023robust}%
  \BibitemOpen
  \bibfield  {author} {\bibinfo {author} {\bibfnamefont {Y.}~\bibnamefont
  {Shapira}}, \bibinfo {author} {\bibfnamefont {S.}~\bibnamefont {Cohen}},
  \bibinfo {author} {\bibfnamefont {N.}~\bibnamefont {Akerman}}, \bibinfo
  {author} {\bibfnamefont {A.}~\bibnamefont {Stern}},\ and\ \bibinfo {author}
  {\bibfnamefont {R.}~\bibnamefont {Ozeri}},\ }\bibfield  {title} {\bibinfo
  {title} {Robust two-qubit gates for trapped ions using spin-dependent
  squeezing},\ }\href@noop {} {\bibfield  {journal} {\bibinfo  {journal}
  {Physical Review Letters}\ }\textbf {\bibinfo {volume} {130}},\ \bibinfo
  {pages} {030602} (\bibinfo {year} {2023})}\BibitemShut {NoStop}%
\bibitem [{\citenamefont {Shapira}\ \emph {et~al.}(2018)\citenamefont
  {Shapira}, \citenamefont {Shaniv}, \citenamefont {Manovitz}, \citenamefont
  {Akerman},\ and\ \citenamefont {Ozeri}}]{shapira2018robust}%
  \BibitemOpen
  \bibfield  {author} {\bibinfo {author} {\bibfnamefont {Y.}~\bibnamefont
  {Shapira}}, \bibinfo {author} {\bibfnamefont {R.}~\bibnamefont {Shaniv}},
  \bibinfo {author} {\bibfnamefont {T.}~\bibnamefont {Manovitz}}, \bibinfo
  {author} {\bibfnamefont {N.}~\bibnamefont {Akerman}},\ and\ \bibinfo {author}
  {\bibfnamefont {R.}~\bibnamefont {Ozeri}},\ }\bibfield  {title} {\bibinfo
  {title} {Robust entanglement gates for trapped-ion qubits},\ }\href@noop {}
  {\bibfield  {journal} {\bibinfo  {journal} {Physical Review Letters}\
  }\textbf {\bibinfo {volume} {121}},\ \bibinfo {pages} {180502} (\bibinfo
  {year} {2018})}\BibitemShut {NoStop}%
\bibitem [{\citenamefont {Wang}\ \emph
  {et~al.}(2020{\natexlab{b}})\citenamefont {Wang}, \citenamefont {Crain},
  \citenamefont {Fang}, \citenamefont {Zhang}, \citenamefont {Huang},
  \citenamefont {Liang}, \citenamefont {Leung}, \citenamefont {Brown},\ and\
  \citenamefont {Kim}}]{wang2020high}%
  \BibitemOpen
  \bibfield  {author} {\bibinfo {author} {\bibfnamefont {Y.}~\bibnamefont
  {Wang}}, \bibinfo {author} {\bibfnamefont {S.}~\bibnamefont {Crain}},
  \bibinfo {author} {\bibfnamefont {C.}~\bibnamefont {Fang}}, \bibinfo {author}
  {\bibfnamefont {B.}~\bibnamefont {Zhang}}, \bibinfo {author} {\bibfnamefont
  {S.}~\bibnamefont {Huang}}, \bibinfo {author} {\bibfnamefont
  {Q.}~\bibnamefont {Liang}}, \bibinfo {author} {\bibfnamefont {P.~H.}\
  \bibnamefont {Leung}}, \bibinfo {author} {\bibfnamefont {K.~R.}\ \bibnamefont
  {Brown}},\ and\ \bibinfo {author} {\bibfnamefont {J.}~\bibnamefont {Kim}},\
  }\bibfield  {title} {\bibinfo {title} {High-fidelity two-qubit gates using a
  microelectromechanical-system-based beam steering system for individual qubit
  addressing},\ }\href@noop {} {\bibfield  {journal} {\bibinfo  {journal}
  {Physical Review Letters}\ }\textbf {\bibinfo {volume} {125}},\ \bibinfo
  {pages} {150505} (\bibinfo {year} {2020}{\natexlab{b}})}\BibitemShut
  {NoStop}%
\bibitem [{\citenamefont {Kang}\ \emph {et~al.}(2021)\citenamefont {Kang},
  \citenamefont {Liang}, \citenamefont {Zhang}, \citenamefont {Huang},
  \citenamefont {Wang}, \citenamefont {Fang}, \citenamefont {Kim},\ and\
  \citenamefont {Brown}}]{kang2021batch}%
  \BibitemOpen
  \bibfield  {author} {\bibinfo {author} {\bibfnamefont {M.}~\bibnamefont
  {Kang}}, \bibinfo {author} {\bibfnamefont {Q.}~\bibnamefont {Liang}},
  \bibinfo {author} {\bibfnamefont {B.}~\bibnamefont {Zhang}}, \bibinfo
  {author} {\bibfnamefont {S.}~\bibnamefont {Huang}}, \bibinfo {author}
  {\bibfnamefont {Y.}~\bibnamefont {Wang}}, \bibinfo {author} {\bibfnamefont
  {C.}~\bibnamefont {Fang}}, \bibinfo {author} {\bibfnamefont {J.}~\bibnamefont
  {Kim}},\ and\ \bibinfo {author} {\bibfnamefont {K.~R.}\ \bibnamefont
  {Brown}},\ }\bibfield  {title} {\bibinfo {title} {Batch optimization of
  frequency-modulated pulses for robust two-qubit gates in ion chains},\
  }\href@noop {} {\bibfield  {journal} {\bibinfo  {journal} {Physical Review
  Applied}\ }\textbf {\bibinfo {volume} {16}},\ \bibinfo {pages} {024039}
  (\bibinfo {year} {2021})}\BibitemShut {NoStop}%
\bibitem [{\citenamefont {Grzesiak}\ \emph {et~al.}(2020)\citenamefont
  {Grzesiak}, \citenamefont {Bl{\"u}mel}, \citenamefont {Wright}, \citenamefont
  {Beck}, \citenamefont {Pisenti}, \citenamefont {Li}, \citenamefont {Chaplin},
  \citenamefont {Amini}, \citenamefont {Debnath}, \citenamefont {Chen} \emph
  {et~al.}}]{grzesiak2020efficient}%
  \BibitemOpen
  \bibfield  {author} {\bibinfo {author} {\bibfnamefont {N.}~\bibnamefont
  {Grzesiak}}, \bibinfo {author} {\bibfnamefont {R.}~\bibnamefont
  {Bl{\"u}mel}}, \bibinfo {author} {\bibfnamefont {K.}~\bibnamefont {Wright}},
  \bibinfo {author} {\bibfnamefont {K.~M.}\ \bibnamefont {Beck}}, \bibinfo
  {author} {\bibfnamefont {N.~C.}\ \bibnamefont {Pisenti}}, \bibinfo {author}
  {\bibfnamefont {M.}~\bibnamefont {Li}}, \bibinfo {author} {\bibfnamefont
  {V.}~\bibnamefont {Chaplin}}, \bibinfo {author} {\bibfnamefont {J.~M.}\
  \bibnamefont {Amini}}, \bibinfo {author} {\bibfnamefont {S.}~\bibnamefont
  {Debnath}}, \bibinfo {author} {\bibfnamefont {J.-S.}\ \bibnamefont {Chen}},
  \emph {et~al.},\ }\bibfield  {title} {\bibinfo {title} {Efficient arbitrary
  simultaneously entangling gates on a trapped-ion quantum computer},\
  }\href@noop {} {\bibfield  {journal} {\bibinfo  {journal} {Nature
  communications}\ }\textbf {\bibinfo {volume} {11}},\ \bibinfo {pages} {2963}
  (\bibinfo {year} {2020})}\BibitemShut {NoStop}%
\bibitem [{\citenamefont {Shizhen}\ \emph {et~al.}(2023)\citenamefont
  {Shizhen}, \citenamefont {Jiayi}, \citenamefont {Donghao},\ and\
  \citenamefont {Zhongxiao}}]{shizhen2023optical}%
  \BibitemOpen
  \bibfield  {author} {\bibinfo {author} {\bibfnamefont {W.}~\bibnamefont
  {Shizhen}}, \bibinfo {author} {\bibfnamefont {C.}~\bibnamefont {Jiayi}},
  \bibinfo {author} {\bibfnamefont {L.}~\bibnamefont {Donghao}},\ and\ \bibinfo
  {author} {\bibfnamefont {X.}~\bibnamefont {Zhongxiao}},\ }\bibfield  {title}
  {\bibinfo {title} {Optical tweezer array with suppressed frequency
  intermodulation},\ }\href@noop {} {\bibfield  {journal} {\bibinfo  {journal}
  {Infrared and Laser Engineering}\ }\textbf {\bibinfo {volume} {52}},\
  \bibinfo {pages} {20230128} (\bibinfo {year} {2023})}\BibitemShut {NoStop}%
\end{thebibliography}%

\setcounter{figure}{0} 
\global\long\def\theequation{S\arabic{equation}}%
\global\long\def\thefigure{S\arabic{figure}}%

\subsection{Methods}

\textbf{The Laser system for Raman operation}--
In the experiment, a 532~nm laser was obtained by frequency doubling
a 1064 nm laser and used to perform stimulated Raman operations. To
generate multiple frequency components, phase modulation was applied
to the 1064 nm seed laser using a fiber electro-optic modulator (EOM)
with a modulation frequency of 13.04~GHz.

\begin{figure*}[htbp]
\centering 
\includegraphics[width=14cm]{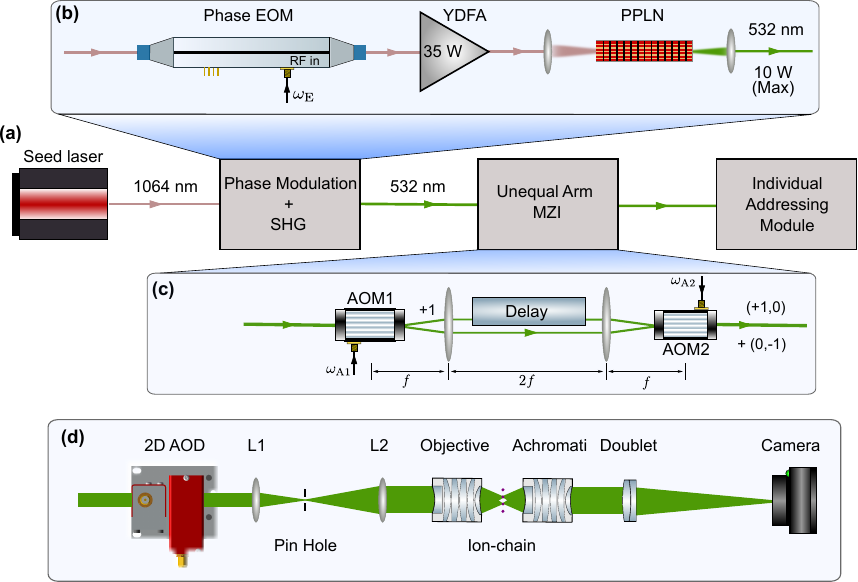} 
\caption{Laser system for Raman transition and individual addressing. (a) Functional
schematic of optical setup. (b) Multi-tone 532 nm laser generation
module. The 1064 nm seed laser is modulated by an EOM to generate
multiple frequency components. Subsequently, a nonlinear crystal with
a bandwidth of approximately 26 GHz converts these components into
a 532 nm laser beam containing multiple frequency components.(c) Unequal-arm
interferometer module. Since driving Raman transitions requires intensity
modulation at 12.64 GHz, we generate an intensity-modulated optical
field through an unequal-arm interferometer composed of a pair of
AOMs. In the experiment, AOM1 is fixed at 200 MHz, while AOM2 is tuned
around a central frequency of 200 MHz to achieve sideband excitation.(d)
Individual addressing module. We implement addressing operations using
a 2D AOD driven by multiple radio-frequency (RF) signals with distinct
frequencies, generating addressable optical spots. After beam expansion,
the Raman laser beam is focused onto ions through a 0.4 NA objective.
For alignment ease, an achromatic objective is used to image both
the ions and 532 nm spots onto a camera simultaneously. (Note: The
369 nm bandpass filter placed before camera is not depicted in the
schematic diagram.)}
\label{fig:figS3}
\end{figure*}

\begin{figure}[htbp]
\centering 
\includegraphics[width=8.5cm]{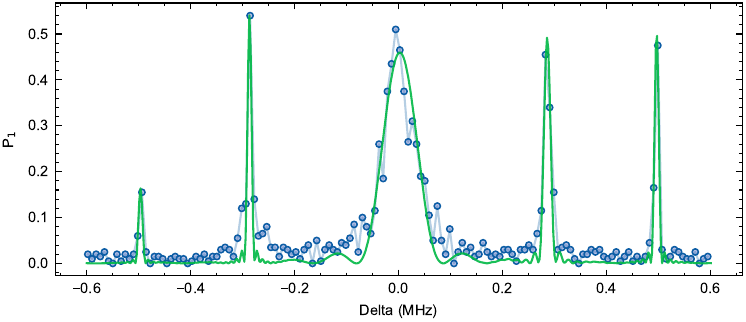} 
\caption{Excitation spectrum of two-ion chain. The ions are positioned at the
center of the light spot. In the experiment, the pulse duration is
fixed at 100~\textmu s, and the detuning of the Raman beam is scanned.}
\label{fig:figS2}
\end{figure}

To generate laser light with an intensity modulation at 12.64 GHz,
we used a compact unbalanced-arm interferometer consisting of a pair
of acousto-optic modulators (AOMs), as shown in Fig.~\ref{fig:figS3}. The radio-frequency
(RF) drive for both AOMs was set to 200 MHz. The $+1^{\text{-st}}$
order diffracted beam from the first AOM (+1,0) and $-1^{\text{-st}}$
order diffracted beam from the second AOM (0,-1) propagated along
the same path, with their interference forming a 12.64 GHz frequency
component that drove the stimulated Raman transitions. Sideband transitions
and MS gate operations were driven by tuning the drive frequency of
the second AOM, as shown in Fig.~\ref{fig:figS2}. To achieve the maximum Rabi frequency,
the diffraction efficiency of the AOMs was adjusted to 50\%. Although
all components with a frequency difference of 12.64 GHz could drive
the Raman transitions, a delay plate of appropriate thickness was
placed in one arm of the interferometer to prevent destructive interference,
which could otherwise lead to a vanishing Rabi frequency.

Here, we present the explicit form of the generated Rabi frequency.
The laser electric field after EOM modulation can be expressed as

\begin{widetext}
\[
E_{1}=\frac{E_{0}}{2}\exp[i(kx-\omega t)]
\sum_{n-\infty}^{\infty}J_{n}(\beta)\times\exp\left\{ i\left[(\delta k)x-\omega_{{\rm {RF}}}t\right]\right\} +{\rm {c.c}}
\]
\end{widetext}
where $\omega_{{\rm {RF}}}$ is the modulation frequency of the EOM,
$\delta k$ denotes the wavenumber shift induced by phase modulation,
$J_{n}(\beta)$ represents the $n$th order Bessel function, and $\beta$
is the modulation depth. In our experiment, $\omega_{{\rm {RF}}}$
was set to 13.04 GHz. And after frequency doubling through the nonlinear
crystal, the expression for the laser electric field can be expressed
as
\begin{widetext}
\[
E_{2}=\eta\frac{E_{0}}{2}\exp[2i(kx-\omega t)]\sum_{n=-\infty}^{\infty}J_{n}(2\beta)\times\exp\left\{ i\left[(\delta k)x-\omega_{{\rm {RF}}}t\right]\right\} +{\rm {c.c}}
\]
\end{widetext} 
where $\eta$ is the harmonic conversion efficiency. After passing
through the unbalanced interferometer, the frequency-doubled laser
generates two distinct electric fields: one produced by the $+1^{\text{-st}}$
order diffraction of the first AOM (+1,0) and the other by the $-1^{\text{-st}}$
order diffraction of the second AOM. We denote these as $E_{+1,0},E_{0,-1}$,
respectively. Assuming that the drive frequency of the first AOM is
$\omega_{\mathrm{AOM}}$ and the second is $\omega_{\mathrm{EOM}}\pm\delta$,
where $\omega_{\mathrm{AOM}}$ was set to 200 MHz, the electric field
can be expressed as
\begin{widetext}
\[
\begin{aligned}E_{+1,0}= & \frac{E_{A1}}{2}\exp[i((2k+\delta k_{A1})x-(2\omega+\omega_{{\rm {AOM}}})t)]\\
 & \times\sum_{n-\infty}^{\infty}J_{n}(2\beta)\times\exp\left\{ in\left[(\delta k)x-\omega_{\mathrm{RF}}t\right]\right\} +{\rm {c.c}}\\
E_{0,-1}= & \frac{E_{A2}}{2}\exp[i(2k-\delta k_{A2})(x+\Delta x)-(2\omega-\omega_{{\rm {AOM}}}-\delta)t)]\\
 & \times\sum_{n-\infty}^{\infty}J_{n}(2\beta)\times\exp\left\{ in\left[(\delta k)(x+\Delta x)-\omega_{\mathrm{RF}}t\right]\right\} +{\rm {c.c}}
\end{aligned}
\]
\end{widetext} 
where $E_{A1}$ and $E_{A2}$ are the amplitudes of the two beams,
$\delta k_{A1}$ and $\delta k_{A2}$ are the wavenumber shifts induced
by the first and second AOMs. The two beams are then combined to form
a single beam with an electric field given by 
\[
E_{3}=E_{+1,0}+E_{0,-1}.
\]
The splitting of the two qubit is defined as $\omega_{{\rm {HF}}}$
= 12.64 GHz, and the Rabi frequency can be calculated as 
\[
\Omega=\frac{\mu_{1}\mu_{2}\langle E_{3}E_{3}^{*}\exp(i\omega_{{\rm {HF}}}t)\rangle}{\hbar^{2}\Delta}.
\]
Considering that $\omega_{{\rm {RF}}}=\omega_{{\rm {HF}}}+2\omega_{{\rm {AOM}}}$,
so the Rabi frequency generated by the modulated laser can be expressed
as 
\[
\Omega=\frac{E_{A1}E_{A2}}{4}\exp\left[i\left(\delta t-\phi(x,\Delta x)\right)\right]f(\delta k\Delta x,\beta)
\]
the expression of $\phi(x,\Delta x)$ and $f(\delta k\Delta x,\beta)$
is given by 
\[
\begin{aligned}\phi(x,\Delta x) & =(\delta k_{A1}+\delta k_{A2})x+\delta k_{A2}\Delta x-\delta kx\\
f(\delta k\Delta x,\beta) & =\sum_{n=-\infty}^{+\infty}J_{n}(2\beta)J_{n-1}(2\beta)\exp(in\delta k\Delta x)
\end{aligned}
\]

\begin{figure}[htbp]
\centering 
\includegraphics[width=8.5cm]{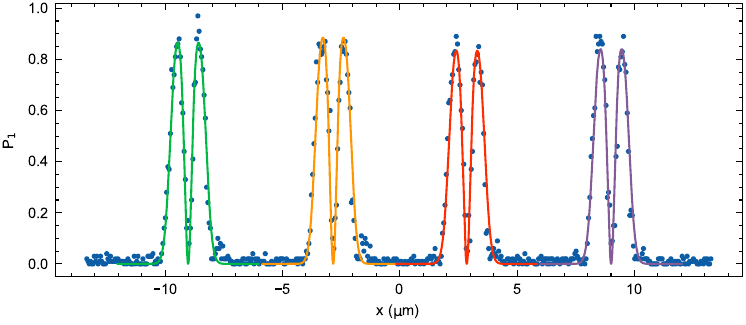} 
\caption{Scanning and fitting the ion positions in the four-ion chain. The
$x$-axis scale is converted from the frequency of the radio-frequency
signal applied to the AOD and calibrated using the axial center-of-mass
(COM) phonon mode frequency and the theoretical spacing of the two-ion
chain, resulting in approximately 5.3 \textmu m/MHz.}
\label{fig:figS1}
\end{figure}

\textbf{Beam alignment and mapping}--
Beam steering and multi-site addressing are accomplished using a two-dimensional
acousto-optic deflector (AOD), with the beam subsequently focused
onto the ion through a 0.4 NA objective lens. Reliable alignment between
the ion and the laser spot is achieved by measuring the $|1\rangle$
state population under fixed-duration Raman laser pulses. The specific
procedure is as follows:

To begin with, a 0.4 NA achromatic objective lens is positioned on
the opposite side of the setup. This lens enables simultaneous imaging
of the 532 nm laser beam and the ion, facilitating the initial alignment
between the laser beam and the ion.At this point, stimulated Raman
transitions can be observed.

Subsequently, the laser spot position is optimized by adjusting the
three-dimensional translation stage of the objective lens to achieve
the highest possible Rabi frequency.

Finally, with the laser pulse duration fixed, the AOD is used to scan
the laser spot position, and the excitation probability of the ion
being driven to the $|1\rangle$ state is measured. By fitting the
data, the ion position can be determined with sub-10 nm precision.
Fig.~\ref{fig:figS1} presents the scanning results for a four-ion
chain.

\textbf{Calibration}--
In the experiment, we calibrate the Rabi frequencies of the red and
blue sidebands as well as the laser intensity on different ions, both
of which are achieved by measuring the carrier Rabi frequency of the
peak.

To calibrate the laser intensity on different ions, the beam alignment
is first completed, after which the AOD is used to simultaneously
generate two laser spots. These spots are then shifted to the positions
where the Rabi frequency reaches its maximum. The peak Rabi frequencies
are measured, and the RF signal strength applied to the AOD is adjusted
to ensure that the peak Rabi frequencies for both ions are equal.

The method for calibrating the intensity of the red and blue sidebands
is similar. Driving signals for the red and blue sidebands $\omega_{\mathrm{AOM}}\pm\delta$
are simultaneously applied to the second AOM, while the EOM drive
frequency is detuned to $\omega_{\mathrm{EOM}}\pm\delta$, respectively.
By measuring the peak carrier Rabi frequency, the signal strength
of the red and blue sidebands can be measured and then calibrated.

\textbf{The profile of Rabi frequency}--
To determine the relative position between the beam center and the
ion, we fitted the beam profile using 
\begin{equation}
\Omega(x)=\Omega_{0}\frac{x}{w_{0}}e^{-2x^{2}/w_{0}^{2}}.
\end{equation}
Here, we provide a further explanation of the origin of this formula.
When a linearly polarized Gaussian beam is tightly focused by a high-NA
objective, the vectorial nature of light becomes significant.We adopt
the Richards-Wolf vectorial diffraction formalism. In spherical coordinates,
the focused electric field (in vacuum) components in the focal region
(z=0) can be written as: 
\begin{widetext}
\begin{equation}
\begin{aligned}E_{x}(x,0,0) & \propto\int_{0}^{\theta_{\mathrm{max}}}\exp\left[-\left(\frac{\sin\theta}{\mathrm{NA}}\right)^{2}\right]\times\sqrt{\cos\theta}\sin\theta(1+\cos\theta)J_{0}(kx\sin\theta)d\theta\\
E_{z}(x,0,0) & \propto\int_{0}^{\theta_{\max}}\exp\left[-\left(\frac{\sin\theta}{\mathrm{NA}}\right)^{2}\right]\times\sqrt{\cos\theta}\sin^{2}\theta J_{1}(kx\sin\theta)d\theta
\end{aligned}
\end{equation}
\end{widetext} 
where $\theta_{\max}=\arcsin(\mathrm{NA})$. These expressions give
the primary electric field components in the $z=0$ plane, near the
focus, along the $x$ direction. $E_{x}$ and $E_{z}$ together form
a two-dimensional elliptical polarization ellipse $\mathbf{E}(x,0,0)=E_{x}(x)\hat{\mathbf{x}}+E_{z}(x)\hat{\mathbf{z}}$.
With $\mathrm{NA}<0.5$, we can get an approximation as 
\begin{equation}
\mathbf{E}\left(x\right)\approx E_{0}\cdot e^{-x^{2}/w_{0}^{2}}\cdot\left(\hat{\mathbf{x}}+i\beta\cdot\frac{x}{w_{0}}\cdot\hat{\mathbf{z}}\right)
\end{equation}
where $w_{0}$ represents the beam waist and $\beta\approx\frac{\pi}{4}\text{NA}$.
For the elliptical polarization $\mathbf{E}=A\hat{\mathbf{x}}+iB\hat{\mathbf{z}}$,
with $A=E_{0}\cdot e^{-x^{2}/w_{0}^{2}}$, $B=A\beta\cdot\frac{x}{w_{0}}$,
we can decompose it into circular polarization basis $\mathbf{E}=E_{+}\hat{R}+E_{-}\hat{L}$,
with $E_{+}=\frac{1}{\sqrt{2}}\left(A+B\right)$ and $E_{-}=\frac{1}{\sqrt{2}}\left(A-B\right)$.

For a qubit encoded in the hyperfine clock transition of $^{2}S_{1/2}$
in $^{171}\mathrm{Yb}^{+}$, with a quantization axis magnetic field
B along the $y$-axis, the circular polarization basis coupling different
transitions from the path in Fig.1(b) (in maintext). As the CG parameter of the transitions
in in form of $(1/\sqrt{3},1/\sqrt{3},-1/\sqrt{3},1/\sqrt{3})$, the
Rabi frequency of Raman laser is 
\begin{equation}
\begin{aligned}\Omega(x) & =\Omega_{+}\Omega_{+}-\Omega_{-}\Omega_{-}\propto E_{+}E_{+}-E_{-}E_{-}\\
 & \propto(A+B)^{2}-(A-B)^{2}=4AB\propto\frac{x}{w_{0}}e^{-2x^{2}/w_{0}^{2}}
\end{aligned}
\end{equation}

\textbf{Energy level shifts}--
The energy level shifts involved in manipulating sidebands for MS gates consist of two components. First, the hyperfine qubit level shift arises from the AC Stark shift induced by the light field, where the hyperfine-split levels exhibit differential AC Stark shifts. Second, the motional frequency shift caused by the optical tweezer (OT) effect modifies the trapping potential. We now examine these mechanisms in detail.

A far-detuned laser induces atomic energy level shifts through the AC Stark effect. Considering a simplified two-level model, the ground state  shift is given by:
\begin{equation}
    U_{T}(\mathbf{r}) = \frac{3\pi c^2}{2\omega_0^3} \frac{\Gamma}{\Delta} I(\mathbf{r}),
\end{equation}
where $\omega_0$ is the atomic transition frequency, $\Gamma$ is the spontaneous decay rate, and $\Delta$ represents the laser detuning from resonance. 

For two hyperfine ground states (L1 and L2) separated by energy $\omega_{\mathrm{HF}}$, the AC Stark shifts $\delta_{\mathrm{L1}}$ and $\delta_{\mathrm{L2}}$ exhibit differential shift
\[
\delta_{\mathrm{L1}} - \delta_{\mathrm{L0}} = U_{T}(\mathbf{r}) \Delta \left( \frac{1}{\Delta-\omega_{\mathrm{HF}}} - \frac{1}{\Delta} \right) 
\approx U_{T}(\mathbf{r}) \frac{\omega_{\mathrm{HF}}}{\Delta},
\]
where the approximation holds for $\Delta \gg \omega_{\mathrm{HF}}$.

For a red-detuned Gaussian beam with intensity profile $I(\rho) = I_0\exp(-\rho^2/w_{0,T}^2)$, the optical tweezer (OT) potential can be approximated as harmonic near the beam center:
\begin{equation}
    U_{T}(\mathbf{\rho}) = U_{T}(0)\exp(-\rho^2/w_{0,T}^2) \approx U_{T}(0)\left(1 - \frac{2\rho^2}{w_{0,T}^2}\right),
\end{equation}
yielding a trapping frequency of:
\begin{equation}
    \omega_T = \sqrt{\frac{4|U_T(0)|}{Mw_{0,T}^2}},
\end{equation}
where $M$ is the atomic mass. 

Considering a single ion trapped in a Paul trap and addressed by the OT, the combined motional frequency becomes:
\[
\omega = \sqrt{\omega_P^2 + \omega_T^2},
\]
where $\omega_P$ is the bare Paul trap frequency. For $\omega_T \ll \omega_P$, this simplifies to:
\[
\omega \approx \omega_P + \frac{\omega_T^2}{2\omega_P}.
\]

\begin{figure}[htbp]
\centering 
\includegraphics[width=8.5cm]{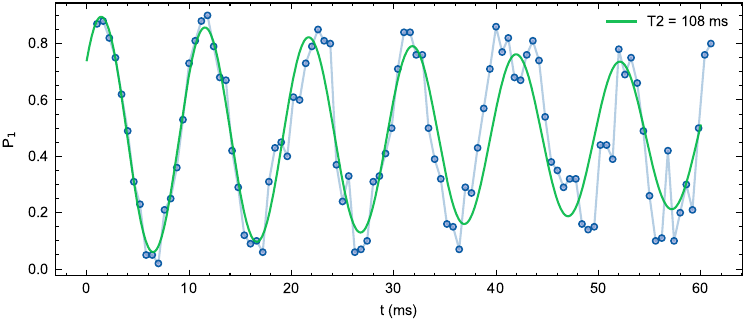} 
\caption{Ramsey oscillation of blue sideband, obtained by scanning the waiting
time between two blue sideband $\pi/2$ pulse, the green curve indicates
a coherent time of T$_{2}$ = 108(35) ms.}
\label{fig:figS5}
\end{figure}

Since we utilized a 532~nm Gaussian beam as the addressing laser
in the experiment, significant optical tweezer effects and differential
AC Stark shifts were observed during the gate operation. The phonon
frequency variation in the STR mode was approximately 10~kHz, and
the differential shift was around 1~kHz. These effects can be measured
and calibrated, while more precise calibration methods will be applied
in our system. Regarding the decoherence caused by fluctuations in
optical intensity, the coherence time of the blue sideband (stretch
mode) exceeds 100~ms, as shown in Fig.~\ref{fig:figS5}, indicating that optical intensity fluctuations
are not the primary factor affecting gate fidelity at this stage.
In the future, we can further optimize the active stabilization of
optical intensity and employ modulated pulse waveforms to reduce the
impact of these disturbances.

\begin{figure*}[htbp]
\centering \includegraphics{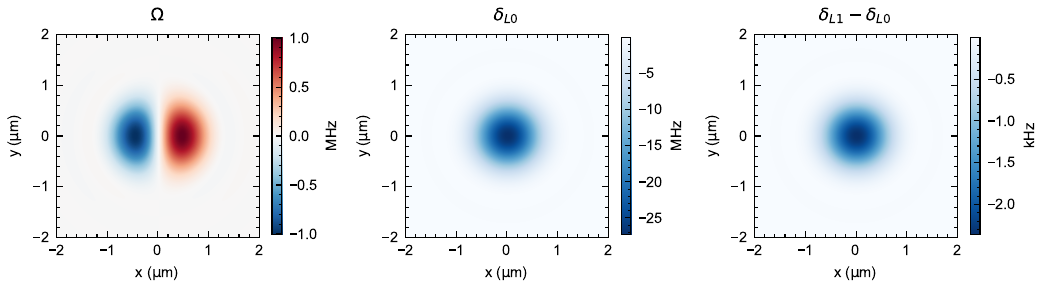} 
\caption{AC Stark shift $\delta_{L,0}$, $\delta_{L,1}$ and the differential
AC Stark shift of the clock states $^{2}S_{1/2}|F=0,\mathrm{m}_{F}=0\rangle$
and $^{2}S_{1/2}|F=1,\mathrm{m}_{F}=0\rangle$ of $^{171}\mathrm{Yb}^{+}$.
The results are simulated under the condition of a 100~mW 532 nm
laser focused by an ideal NA=0.3 objective lens.}
\label{fig:figS6}
\end{figure*}

Since the distribution of the focused light field can be simulated
using the Debye-Wolf vector diffraction theory, the AC Stark shift
can be calculated based on this distribution.

In Fig.~\ref{fig:figS6}, we illustrate the energy level shifts of $|0\rangle$ and
$|1\rangle$ induced by a 100~mW 532~nm laser focused by an ideal
objective lens of NA=0.3, where the beam waist matches the experimentally
measured value. In this work, the AC Stark shift induced by the Raman
beam is approximately 7.2~MHz, based on fluorescence count measurements
during cooling. According to the simulation results, this corresponds
to an optical tweezer effect of 100 kHz and a differential shift of
0.62 kHz, which are in agreement with the experimental results.

\end{document}